\documentclass{article} 
\usepackage{iclr2027_conference,times}

\usepackage{amsmath,amsfonts,bm}

\def\eqref#1{equation~\ref{#1}}

\def\1{\bm{1}}

\DeclareMathAlphabet{\mathsfit}{\encodingdefault}{\sfdefault}{m}{sl}
\SetMathAlphabet{\mathsfit}{bold}{\encodingdefault}{\sfdefault}{bx}{n}

\usepackage{graphicx}
\usepackage{url}
\usepackage{bbm}
\usepackage{orcidlink}
\usepackage{wrapfig}
\usepackage{siunitx}
\usepackage{algorithm}
\usepackage[noend]{algpseudocode}
\usepackage{mathrsfs}
\usepackage{tabularx}
\usepackage{booktabs}
\usepackage{multirow}

\usepackage{hyperref}
\hypersetup{
    colorlinks=true,
    linkcolor=[rgb]{0.87,0.18,0.22},
    citecolor=[rgb]{0.0,0.60,0.33},
    urlcolor=[rgb]{0.20,0.30,0.54}
}

\usepackage{xcolor}

\preprint

\title{Neural-Network Solutions to Real-Space Charge Density and Generalization}

\author{%
Yuxuan Zeng\orcidlink{0009-0004-8954-4377}$^\text{1,2}$, Tong Xie$^\text{3,4}$\thanks{Corresponding author.}, Taoyuze Lv\orcidlink{0000-0002-8281-5048}$^\text{1,2}$\thanks{Corresponding author.}, Zhicheng Zhong\orcidlink{0000-0003-1507-4814
}$^\text{1,2}$\thanks{Corresponding author.}\\
\texttt{\href{mailto:tong.xie@unsw.edu.au}{tong.xie@unsw.edu.au}, \{\href{mailto:taoyuze.lv@ustc.edu.cn}{taoyuze.lv}, \href{mailto:zczhong@ustc.edu.cn}{zczhong}\}@ustc.edu.cn}\\
$^\text{1}$School of Artificial Intelligence \& Data Science, University of Science and Technology of China\\
$^\text{2}$Suzhou Institute for Advanced Research, University of Science and Technology of China\\
$^\text{3}$University of New South Wales, NSW, Sydney, Australia\\
$^\text{4}$Green Dynamics
}

\begin{document}

\maketitle

\begin{abstract}
The Hohenberg-Kohn theorem establishes that, in principle, the ground state (GS) charge density contains all GS information of a many-electron system, such that all GS observables can be expressed as functionals of the GS charge density. Conventional Kohn-Sham density functional theory requires iterative solution of the self-consistent-field (SCF) equations at substantial computational cost, motivating the development of deep learning surrogates for electronic structure calculations and, in turn, accelerating computer-aided materials design. Here, we propose \textbf{AIDEN}, an \underline{A}tomic-\underline{I}nteraction \underline{D}ensity \underline{E}quivariant \underline{N}etwork for solving real-space charge density. AIDEN separates the element-dependent one-center density from environment-induced density redistribution and represents the latter through complementary atom- and edge-centered tensor correlations. A continuous low-rank Gaussian decoder then reconstructs the density at arbitrary spatial coordinates while reusing atomic encodings independently of the evaluation grid. AIDEN achieves state-of-the-art accuracy on periodic crystal benchmarks while remaining competitive for molecular systems, and further demonstrates zero-shot transferability across several structurally distinct out-of-distribution case studies. Furthermore, AIDEN provides substantially faster inference than both baseline models and full SCF calculations, enabling efficient charge density reconstruction for large-scale electronic structure calculations.
\end{abstract}

\section{Introduction}
\label{sec:intro}

The exploration of the vast materials space is rapidly shifting from conventional trial-and-error experiments and direct first-principles calculations toward data-driven discovery, with deep learning playing an increasingly important role. The data underlying this paradigm are often generated from high-fidelity Kohn-Sham density functional theory (KS-DFT) calculations~\citep{RevModPhys.87.897} or molecular dynamics (MD) simulations~\citep{md-rev}, which can be performed systematically at scale and generally provide greater internal consistency than experimental measurements. Deep-learning approaches to DFT calculations commonly target the charge density~\citep{eacnet}, Hamiltonian~\citep{tang2024deep}, and density matrix~\citep{wl9w-8g8r}, which are connected through the self-consistent field (SCF) cycle~\citep{scf-ref}:
\begin{align}
    \rho(\mathbf{r}) \to H_{\rm KS}[\rho] \to \left\{\phi_{n\mathbf{k}},\epsilon_{n\mathbf{k}}\right\} \to \mathbf{D} \to \rho(\mathbf{r}).
    \label{eq:scf}
\end{align}
Here, $\rho(\mathbf{r})$ denotes the real-space charge density, $H_{\rm KS}[\rho]$ the KS Hamiltonian constructed from the density-dependent effective potential, and $\mathbf{D}$ the one-particle density matrix constructed from the KS orbitals and their occupations. Compared with learning the Hamiltonian or density matrix, charge density prediction offers several advantages:
\begin{itemize}
    \item[1)] \textbf{Weaker dependence on atomic-orbital basis representations.} Changing the atomic-orbital basis alters the matrix representations of both the Hamiltonian and density matrix~\citep{deeph-r}, whereas $\rho(\mathbf{r})$ is intrinsically a real-space scalar field independent of orbital indexing, facilitating a more unified representation across different basis sets.

    \item[2)] \textbf{Direct physical interpretability.} The real-space charge density provides direct access to electronic-structure characteristics such as charge transfer~\citep{poli2020charge} and chemical bonding~\citep{charge-bond}, offering a microscopic basis for interpreting material properties.

    \item[3)] \textbf{Convenient three-dimensional (3D) representation.} When discretized on a real-space grid, $\rho(\mathbf{r})$ forms a regular 3D tensor and can serve as a complementary electronic-structure descriptor for downstream representation learning~\citep{shuang2026edditphysicsguideddiffusionpretraining}.
\end{itemize}

To this end, we propose an \underline{A}tomic-\underline{I}nteraction \underline{D}ensity \underline{E}quivariant \underline{N}etwork (AIDEN) to efficiently predict real-space GS charge densities directly from atomic configurations. Inspired by the superposition of atomic densities (SAD)~\citep{sad}, AIDEN decomposes the total charge density into an atom-centered term $\rho_{\rm init}(\mathbf{r})$ and an environment-dependent term $\rho_{\rm env}(\mathbf{r})$. The former serves as a learnable element-dependent one-center baseline for the dominant local density distribution, whereas the latter captures environment-induced density variations through higher-order many-body features. In the encoder, AIDEN first employs Cartesian atomic cluster expansion (ACE)~\citep{tace} to encode higher-order geometric and angular information of local atomic environments, followed by tensor edge cluster expansion (TECE)~\citep{tece} to equivariantly model and adaptively aggregate direction-dependent neighbor interactions, providing expressive representations of complex local electronic environments. The decoder then expands the atomic features using Gaussian-type orbitals (GTOs)~\citep{gto} and reconstructs a continuous real-space charge-density field. In addition, the computationally expensive equivariant atomic encoding in AIDEN is performed only once and reused across all spatial query points, yielding impressively faster inference than conventional probe-based methods. Together, these designs provide AIDEN with outstanding learning capacity and substantially improved efficiency, while enabling its extension to larger-scale out-of-distribution (OOD) systems.

\section{Related works}
\textbf{Equivariant GNNs.}
Conventional atomistic GNNs mainly learn rotation- and translation-invariant scalar representations~\citep{cgcnn}, whereas equivariant GNNs enforce $f(D_{\mathcal X}(g)x)=D_{\mathcal Y}(g)f(x)$ to preserve geometric transformation laws. Tensor Field Networks~\citep{Zaccone2026} and e3nn~\citep{e3nn} established E(3)-equivariant representations based on SO(3) irreducible representations and spherical harmonics. NequIP~\citep{nequip} demonstrated their effectiveness for atomistic modeling, while MACE~\citep{mace} incorporated atomic cluster expansion to capture higher-order many-body interactions. More recent methods improve efficiency through edge-aligned SO(2) operations, including eSCN~\citep{pmlr-v202-passaro23a} and TECE~\citep{tece}, the latter further introducing edge cluster expansion and radial rotary attention.

\textbf{Charge density learning.}
Recent methods predict real-space charge densities using spatial queries, regular grids, or atom-centered representations. DeepDFT~\citep{jorgensen2022equivariant} predicts densities at arbitrary probe points from local atomic environments, while Deep Charge~\citep{deepcharge} learns symmetry-preserving local representations with good data efficiency. ChargE3Net~\citep{koker2024higher} extends probe-based prediction with higher-order E(3)-equivariant features and scales to more than $10^5$ crystals. \cite{li2025image} instead reconstructs molecular densities on regular 3D grids, whereas EAC-Net~\citep{eacnet} represents the density as a sum of symmetry-consistent atomic contributions. NeuralSCF~\citep{song2026neural} further learns the Kohn-Sham density map through neural self-consistent iterations.

\section{Problem statement}
\label{sec:problem}

Consider a unit cell $\Omega=\{\mathbf u\mathbf A\mid\mathbf u\in[0,1)^3\}$ with lattice vectors arranged by rows in $\mathbf A\in\mathbb R^{3\times3}$. A periodic structure is written as $\mathcal X=(\mathbf A,\{Z_i,\mathbf s_i\}_{i=1}^{N_{\rm a}})$, where $Z_i$ and $\mathbf s_i\in[0,1)^3$ are the atomic number and fractional coordinate of atom $i$, respectively, and $\mathbf R_i=\mathbf s_i\mathbf A$ is its Cartesian position. Its periodic images lie at $\mathbf R_i+\mathbf n\mathbf A$ for $\mathbf n\in\mathbb Z^3$. For a cell containing $N_{\rm e}$ electrons, the physical GS density belongs to
\begin{align}
\mathcal D_{N_{\rm e}}=\left\{\rho:\Omega\to\mathbb R_+\ \middle|\ \int_\Omega\rho(\mathbf r)\,{\rm d}^3\mathbf r=N_{\rm e}\right\},\quad \rho_{\mathcal X}(\mathbf r)=N_{\rm e}\int \prod_{a=2}^{N_{\rm e}}d^3\mathbf r_a \left|\Psi(\mathbf r,\mathbf r_2,\dots,\mathbf r_{N_{\rm e}})\right|^2,
\label{eq:physical-density}
\end{align}
and KS-DFT obtains $\rho_{\mathcal X}^{\star}=\arg\min_{\rho\in\mathcal D_{N_{\rm e}}}E_{\mathcal X}[\rho]$. Plane-wave calculations~\citep{pwdft} represent the real-space density numerically on a regular fast Fourier transform (FFT) grid~\citep{3dfft}, $\mathcal R_{\mathcal X}=\{\mathbf r_g\}_{g=1}^{N_{\rm g}}$. Below, $\rho$ denotes the real-space FFT-grid density used as the learning target, and $N_{\rm e}$ denotes the corresponding electron count, with $N_{\rm e}\simeq|\Omega|N_{\rm g}^{-1}\sum_g\rho_{\mathcal X}(\mathbf r_g)$. AIDEN learns a continuous scalar field $\widehat\rho_{\boldsymbol\theta}(\mathbf r;\mathcal X)$. For a Euclidean transformation $g=(\mathbf Q,\mathbf t)$ with $\mathbf Q\in{\rm O}(3)$ and $g\mathbf r=\mathbf r\mathbf Q^\top+\mathbf t$, the target covariance and lattice periodicity are $\widehat\rho_{\boldsymbol\theta}(g\mathbf r;g\mathcal X)=\widehat\rho_{\boldsymbol\theta}(\mathbf r;\mathcal X)$ and $\widehat\rho_{\boldsymbol\theta}(\mathbf r+\mathbf n\mathbf A;\mathcal X)=\widehat\rho_{\boldsymbol\theta}(\mathbf r;\mathcal X)$. Given training samples $\{\mathcal X_n\}_{n=1}^{N_{\rm d}}$ with full FFT grids $\mathcal R_{\mathcal X_n}=\{\mathbf r_{ng}\}_{g=1}^{N_{\rm g}^{(n)}}$, the model minimizes the grid-averaged absolute deviation to determine the optimal model parameters $\theta^{\star}$,
\begin{align}
\boldsymbol\theta^\star={\arg\min}_{\boldsymbol\theta}\frac{1}{N_{\rm d}}\sum_{n=1}^{N_{\rm d}}\frac{1}{N_{\rm g}^{(n)}}\sum_{g=1}^{N_{\rm g}^{(n)}}\left|\widehat\rho_{\boldsymbol\theta}(\mathbf r_{ng};\mathcal X_n)-\rho_{\mathcal X_n}(\mathbf r_{ng})\right|.
\label{eq:training-objective}
\end{align}

\section{Model architecture}
\label{sec:model}

\begin{figure}[t]
    \centering
    \includegraphics[width=\linewidth]{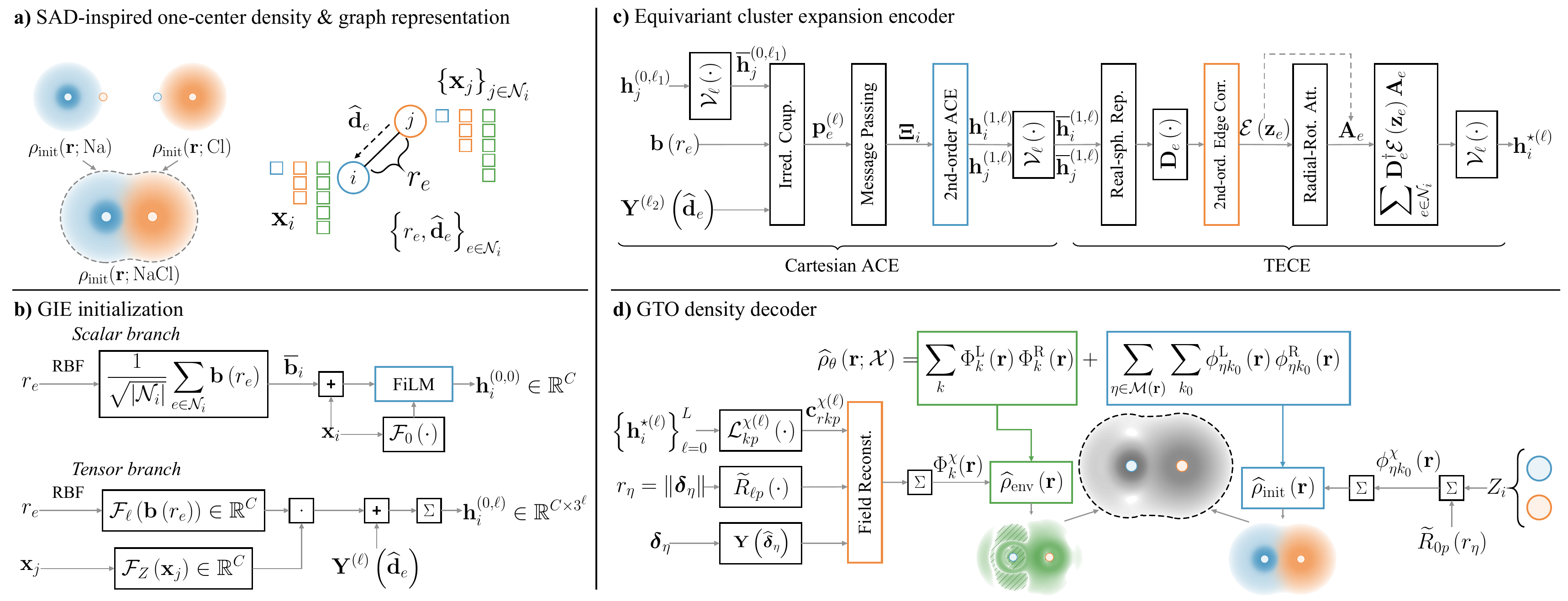}
    \caption{\textbf{Overview of AIDEN.} (a) Using a simple Na-Cl atom pair as an example, the initial local charge density $\rho_{\rm init}(\mathbf{r};\text{NaCl})$ can be decomposed into separate contributions from the two atomic species, while the atomic system is represented as a periodic graph containing elemental embeddings, interatomic distances, and edge-direction information. (b) The geometry-informed embedding (GIE) module initializes scalar and higher-order equivariant atomic features through coordinated scalar and tensor branches. (c) In the encoder, Cartesian ACE constructs many-body correlations after neighborhood aggregation, whereas TECE builds higher-order source-target correlations in edge-aligned local coordinate frames and further modulates the correlated information through radial rotary attention (RRA). (d) In the decoder, the charge density is composed of a local term $\rho_{\rm init}(\mathbf{r})$ and an environment term $\rho_{\rm env}(\mathbf{r})$; the equivariant atomic representations are mapped to local GTO coefficients and can be efficiently evaluated at arbitrary spatial coordinates.}
    \label{fig:model-arch}
    
\end{figure}

\subsection{Periodic geometry and equivariant initialization}
\label{sec:encoder}

For every target atom $i$, the periodic atomic graph contains directed edges $e=(j,\mathbf n\!\to i)$ satisfying
\begin{align}
\mathbf d_e=\mathbf R_i-(\mathbf R_j+\mathbf n\mathbf A),\quad r_e=\|\mathbf d_e\|_2,\quad \widehat{\mathbf d}_e=\mathbf d_e/r_e,\quad 0<r_e<r_{\rm at}.
\label{eq:atomic-edge}
\end{align}
If more than $N_{\rm nbr}$ images enter the cutoff sphere, only the nearest $N_{\rm nbr}$ are retained for that target. We write $\mathcal N_i$ for the resulting incoming edges. Each edge carries a radial vector $\mathbf b(r_e)\in\mathbb R^{N_{\rm B}}$ and an irreducible rank-$\ell$ Cartesian harmonic $\mathbf Y^{(\ell)}(\widehat{\mathbf d}_e)$. Their exact construction is given in Appx.~\ref{app:graph-basis}. 

The element input is $\mathbf x_i={\rm Norm}[\mathbf o(Z_i)\oplus\mathbf q(Z_i)]\in[0,1]^{133}$, where $\mathbf o(Z_i)\in\{0,1\}^{118}$ is a one-hot element vector, $\mathbf q(Z_i)\in\mathbb R^{15}$ contains the elemental descriptors, $\oplus$ denotes concatenation, and ${\rm Norm}$ is feature-wise min-max normalization over the 118 elements. GIE initializes the scalar and higher-order tensor fields directly from the one-hop atomic environment. We first summarize the radial environment of atom $i$ by
\begin{align}
\overline{\mathbf b}_i&=\frac{1}{\sqrt{|\mathcal N_i|}}\sum_{e\in\mathcal N_i}\mathbf b\left(r_e\right),\quad
\mathbf h_i^{\left(0,0\right)}=\mathcal F_{\rm s}\left[\mathcal F_0\left(\mathbf x_i\right),\mathbf x_i\oplus\overline{\mathbf b}_i\right],\label{eq:gie-scalar}\\
\mathbf h_i^{\left(0,\ell\right)}&=\frac{1}{\sqrt{|\mathcal N_i|}}\mathcal{P}_{\ell}^{\rm irr}\left\{\left[\sum_{e\in \mathcal N_i}\mathcal F_{\ell}\left[\mathbf b\left(r_e\right)\right]\odot\mathcal F_{\rm Z}\left(\mathbf x_j\right)\right]\otimes\mathbf Y^{\left(\ell\right)}\left(\widehat{\mathbf d}_e\right)\right\},\quad 1\le\ell\le L.
\label{eq:gie-compact}
\end{align}
For the scalar field, $\mathcal F_0\left(\mathbf x_i\right)$ provides the elemental seed of the central atom, while $\mathcal F_{\rm s}\left(\cdot\right)$ applies a FiLM-style environment-conditioned~\citep{perez2018film} scale and shift determined from $\mathbf x_i\oplus\overline{\mathbf b}_i$. For $\ell\ge1$, $\mathcal F_{\ell}[\mathbf b\left(r_e\right)]\in\mathbb R^C$ assigns order-specific radial amplitudes, controlling how strongly a neighbor at distance $r_e$ contributes to each channel, whereas $\mathcal F_{\rm Z}\left(\mathbf x_j\right)\in\mathbb R^C$ supplies elemental amplitudes that modulate these contributions according to the chemical identity of the source atom. Their channel-wise product therefore determines the magnitude of each edge contribution, while the analytic harmonic $\mathbf Y^{\left(\ell\right)}\left(\widehat{\mathbf d}_e\right)$ fixes its angular dependence and transformation law. Here $\odot$ acts over the $C$ scalar channels, and $\otimes$ broadcasts each channel weight over the Cartesian components of the rank-$\ell$ harmonic. The projection $\mathcal P_{\ell}^{\rm irr}$ keeps the accumulated tensor in the symmetric traceless rank-$\ell$ subspace. Consequently, $\mathbf h_i^{\left(0,\ell\right)}\in\mathbb R^{C\times3^\ell}$ forms an equivariant Cartesian field in which the learned maps determine the channel-wise strength of neighboring contributions without independently parameterizing individual Cartesian components, thereby preserving rotational equivariance.

\subsection{Equivariant cluster expansion encoder}
\label{sec:cluster-encoder}

Before each interaction, every angular order is normalized as $\overline{\mathbf h}_i^{(t,\ell)}=\mathcal V_{\ell}[\mathbf h_i^{(t,\ell)}]$ (Appx.~\ref{app:ace-detail}). In our case, one Cartesian ACE interaction forms higher-order correlations after neighborhood aggregation, describing collective atom-centered coordination. One subsequent TECE interaction forms source-target correlations within each edge-aligned frame before aggregation, retaining directional information along individual interatomic axes. These are complementary geometric inductive biases for density redistribution.

\subsubsection{Cartesian atomic cluster expansion}

The Cartesian ACE follows the irreducible tensor construction of TACE~\citep{tace}. Let $\mathcal T_L$ be the admissible angular paths defined in Eq.~\ref{eq:ace-paths}, and let $\mathcal C_{\ell_1,\ell_2}^{\ell}$ be the irreducible Cartesian coupling defined in Eq.~\ref{eq:cart-coupling}. The edge-resolved one-particle tensor is
\begin{align}
\mathbf p_e^{(\ell)}=\sum_{(\ell_1,\ell_2,\ell)\in\mathcal T_L}\mathcal C_{\ell_1,\ell_2}^{\ell}\left(\mathbf w_{e,\ell_1\ell_2\ell}\odot\mathcal L_{\rm s}^{(\ell_1)}\overline{\mathbf h}_j^{(0,\ell_1)},\mathbf Y^{(\ell_2)}(\widehat{\mathbf d}_e)\right),\quad \mathbf w_{e,\ell_1\ell_2\ell}=\mathcal F_{\rm ACE}^{\ell_1\ell_2\ell}\left[\mathbf b(r_e)\right],
\label{eq:ace-edge-field}
\end{align}
where $\mathcal L_{\rm s}^{(\ell)}$ mixes channels at fixed $\ell$ and $\mathcal F_{\rm ACE}^{\ell_1\ell_2\ell}$ produces a $C$-component radial weight. The invariant neighbor coefficient $a_e$ and the explicit construction of the order-$\nu$ Cartesian correlation polynomial $\mathcal B_{\nu}^{(\ell)}$ are detailed in Appx.~\ref{app:ace-detail}. Define the aggregated atomic field $\boldsymbol\Xi_i$ as
\begin{align}
\boldsymbol\Xi_i&=\mathcal L_{\rm self}\overline{\mathbf h}_i^{(0)}+\mathcal L_{\rm msg}\sum_{e\in\mathcal N_i}a_e\mathbf p_e,\notag\\
\mathbf h_i^{(1,\ell)}&=\overline{\mathbf h}_i^{(0,\ell)}+\mathcal L_{\rm out}^{(\ell)}\mathcal B_{\nu}^{(\ell)}\left\{\left[\mathbf 1_C+\sigma_0\mathcal F_{\rm g}\left(\boldsymbol\Xi_i^{(0)}\right)\right]\odot\boldsymbol\Xi_i\right\},
\label{eq:ace-compact-update}
\end{align}
where $\sigma_0(x)=(1+{\rm e}^{-x})^{-1}$ is the sigmoid function, and $\mathcal F_{\rm g}$ maps the invariant sector $\boldsymbol\Xi_i^{(0)}$ to channel-wise gating coefficients. The symbol $\mathbf1_C$ is the all-ones channel vector and is broadcast over every angular component. Bold symbols without an angular superscript denote the direct sum over $\ell=0,\dots,L$, and every $\mathcal L$ acts only on channels of equal angular order.

\subsubsection{Tensor edge cluster expansion and radial rotary attention}

We first transform the irreducible Cartesian features into a real-spherical representation and then rotate them into an edge-aligned local frame. Let $\mathcal U$ be the fixed orthogonal map from irreducible Cartesian tensors to real-spherical components, and let $\mathbf D_e$ align the local $y$ axis with $\widehat{\mathbf d}_e$. With $M$ denoting the maximum retained magnetic order, restricting to $|m|\le M$ gives $D_M=(L+1)+2\sum_{m=1}^{M}(L+1-m)$ real components. The radial operator $\boldsymbol\Omega_e[\mathbf b(r_e)]$ and the second-order local edge correlator $\mathcal E$ are detailed in Appx.~\ref{app:tece-detail}. The complete TECE update is the single composition
\begin{align}
\mathbf h_i^{(2)}&=\frac{1}{\sqrt2}\left\{\overline{\mathbf h}_i^{(1)}+\mathcal U^\dagger\sum_{e\in\mathcal N_i}\mathbf D_e^\dagger\,\mathcal E\left[\boldsymbol\Omega_e[\mathbf b(r_e)]\odot\left(\mathbf D_e\mathcal U\overline{\mathbf h}_j^{(1)}\oplus\mathbf D_e\mathcal U\overline{\mathbf h}_i^{(1)}\right)\right]\mathbf A_e\right\},\label{eq:tece-compact-update}\\
\mathbf h_i^{\star(\ell)}&=\mathcal V_{\ell}[\mathbf h_i^{(2,\ell)}].\label{eq:tece-output-normalization}
\end{align}
Here the inverse transforms $\mathbf D_e^\dagger$ and $\mathcal U^\dagger$ return the correlated edge features to the global Cartesian representation. The product $\odot$ in Eq.~\ref{eq:tece-compact-update} acts elementwise over the compact components and channels. The $H$ attention heads, each containing $C_h=C_{\rm e}/H$ edge channels, act through $\mathbf A_e={\rm diag}(a_{e1}\mathbf I_{C_h},\dots,a_{eH}\mathbf I_{C_h})$. The query $\mathbf q_{e,\ell mh}\in\mathbb C^{C_h}$ and key $\mathbf k_{e,\ell mh}\in\mathbb C^{C_h}$ are channel projections of the unmodulated local target and source tensors, respectively; their $m=0$ components are real. For $\mathbf u,\mathbf v\in\mathbb C^{C_h}$, the head-wise Hermitian contraction is $\langle\mathbf u,\mathbf v\rangle_{\rm ch}=\sum_{c=1}^{C_h}u_c^*v_c$. Radial rotary attention assigns
\begin{align}
\xi_{eh}=\frac{\tau_h}{\sqrt{D_MC_h}}\left(\sum_{\ell=0}^{L}\left\langle\mathbf q_{e,\ell0h},\mathbf k_{e,\ell0h}\right\rangle_{\rm ch}+\sum_{m=1}^{M}\sum_{\ell=m}^{L}{\rm Re}\left\langle\mathbf q_{e,\ell mh},{\rm e}^{{\rm i}m\varphi_{eh}}\mathbf k_{e,\ell mh}\right\rangle_{\rm ch}\right)+\beta_{eh},
\label{eq:rra-score}
\end{align}
where $\beta_{eh}\in\mathbb R$ and $\varphi_{eh}\in(-\pi,\pi)$ are radial bias and phase, while $\tau_h>0$ is a learned inverse temperature controlling the sharpness of the attention distribution for head $h$. The normalized coefficient $a_{eh}$ is a cutoff-weighted softmax over incoming edges, as detailed in Appx.~\ref{app:tece-detail}. The score consists only of invariant inner products of equal local frequencies; the radial phase adjusts their relative alignment without changing the SO(2) transformation law.

\subsection{Continuous density decoder}
\label{sec:decoder}

The decoder maps $\mathbf h_i^{\star(\ell)}$ to rank-$\ell$ coefficient tensors $\mathbf c_{ikp}^{\chi(\ell)}=\mathcal L_{kp}^{\chi(\ell)}\mathbf h_i^{\star(\ell)}+\delta_{\ell0}\mu_{kp}^{\chi}$, where $\chi\in\{\mathrm L,\mathrm R\}$, $k=1,\dots,K$, and $p=1,\dots,N_{\rm G}$. Here, $\mathcal L_{kp}^{\chi(\ell)}$ is an equivariant channel map, $\delta_{\ell0}$ is the Kronecker delta, and $\mu_{kp}^{\chi}$ is a scalar bias restricted to $\ell=0$. The quantities $K$ and $N_{\rm G}$ denote the environmental rank and the number of Gaussian radial functions, respectively. This Cartesian formulation is equivalent to expressing both the coefficients and harmonics in the same orthogonal real-spherical basis. For a query position $\mathbf r$, we define the supported periodic atom images as
\begin{align}
\mathcal M(\mathbf r)=\left\{\eta=(i,\mathbf n)\ \middle|\ \boldsymbol\delta_\eta=\mathbf r-(\mathbf R_i+\mathbf n\mathbf A),\ r_\eta=\|\boldsymbol\delta_\eta\|_2<r_{\rm orb}\right\},\quad \widehat{\boldsymbol\delta}_\eta=\begin{cases}\boldsymbol\delta_\eta/r_\eta,&r_\eta>0,\\ \mathbf0,&r_\eta=0.\end{cases}
\label{eq:query-images}
\end{align}

At an atomic center, we adopt the continuous convention $\mathbf Y^{(0)}(\mathbf0)=1$ and $\mathbf Y^{(\ell)}(\mathbf0)=\mathbf0$ for $\ell>0$. The radial functions $\widetilde R_{\ell p}(r)$ follow a corrected even-tempered Gaussian construction, while $\upsilon_{Z_i k_0p}^{\chi}$ denotes an element-indexed learnable coefficient for one-center rank $k_0=1,\dots,K_0$. Their definitions and the detailed Gaussian basis construction are given in Appx.~\ref{app:decoder-detail}. Let $\langle\cdot,\cdot\rangle_{\rm F}$ denote contraction over all Cartesian tensor indices. The element-only contribution associated with image $\eta$ and the environment-dependent field at a query position are then
\begin{align}
\phi_{\eta k_0}^{\chi}(\mathbf r)=\sum_{p}\upsilon_{Z_i k_0p}^{\chi}\widetilde R_{0p}(r_\eta),\quad
\Phi_k^{\chi}(\mathbf r)=\sum_{\eta\in\mathcal M(\mathbf r)}\sum_{\ell,p}\frac{\widetilde R_{\ell p}(r_\eta)}{3^{\ell/2}}\left\langle\mathbf c_{ikp}^{\chi(\ell)},\mathbf Y^{(\ell)}\left(\widehat{\boldsymbol\delta}_\eta\right)\right\rangle_{\rm F}.
\label{eq:continuous-fields}
\end{align}

The continuous density is reconstructed as
\begin{align}
\widehat\rho_{\boldsymbol\theta}\left(\mathbf r;\mathcal X\right)=\underbrace{\sum_{\eta\in\mathcal M\left(\mathbf r\right)}\sum_{k_0}\phi_{\eta k_0}^{\mathrm L}\left(\mathbf r\right)\phi_{\eta k_0}^{\mathrm R}\left(\mathbf r\right)}_{\rho_{\rm init}\left(\mathbf r\right)}+\underbrace{\sum_{k}\Phi_k^{\mathrm L}\left(\mathbf r\right)\Phi_k^{\mathrm R}\left(\mathbf r\right)}_{\rho_{\rm env}\left(\mathbf r\right)}.
\label{eq:density-master}
\end{align}

Here $\rho_{\rm init}$ is an element-dependent one-center density independent of the encoded environment, whereas $\rho_{\rm env}$ depends on $\mathbf h_i^{\star(\ell)}$. In $\rho_{\rm env}$, the left and right fields are accumulated over $\mathcal M(\mathbf r)$ before multiplication, naturally introducing cross terms between distinct atom images without explicit pair enumeration. Both terms are expanded in atom-centered GTOs. In our case, we first perform a short pretraining stage dedicated to initializing the representation network for $\rho_{\rm init}$. During subsequent full training, $\rho_{\rm init}$ and $\rho_{\rm env}$ are jointly optimized. The complete forward process is provided in Appx.~\ref{app:forward}.

\section{Experiments}
\label{sec:exp}

\subsection{Dataset settings}

Although AIDEN is primarily designed for periodic crystals, we further evaluate its applicability to both crystalline and molecular systems. For crystals, we use the ECD dataset~\citep{chen2025ecd}, which contains 140,646 inorganic structures spanning 94 elements. The reference charge densities are mainly computed with VASP~\citep{VASPKIT} using PAW-PBE~\citep{pbe}, with PBE+U~\citep{pbe+u} for selected strongly correlated systems and a plane-wave cutoff of 520~eV; an additional 7,147 samples are recalculated with HSE06. We train AIDEN from scratch on the PBE subset and fine-tune it on the smaller HSE subset. For molecular systems, we use the QM9 charge density dataset~\citep{jorgensen2022equivariant}, containing 133,885 molecules composed of H, C, N, O, and F, with densities computed using VASP and PAW-PBE at a 400~eV cutoff and $\Gamma$-point sampling. These datasets provide complementary periodic and molecular regimes for evaluating generalization across chemical spaces, structural scales, and boundary conditions. For OOD evaluation, we additionally consider liquid water, disordered AlMg, amorphous silicon (a-Si), and twisted bilayer graphene (TBG). The first three systems contain 192, 108, and 64 atoms, respectively, while the largest of three TBG structures contains $N=148$ atoms.

\subsection{Evaluation on benchmark datasets}
\label{sec:bench}

Real-space charge density is inherently a continuous physical quantity. For computational convenience, a common coarse-graining strategy is to sample it on a 3D grid, which typically contains millions to tens of millions of grid points. In practice, however, only a tiny fraction of the available density grid can be sufficient for accurate model fitting, particularly when non-uniform or targeted sampling strategies are employed~\citep{jorgensen2022equivariant,focassio2023linear}. In our case, we uniformly sample 5,000 grid points from each structure for training and quantify the error using the normalized mean absolute error (NMAE),
\begin{align}
    \varepsilon_{\rho}:=\frac{1}{N}\sum_{n=1}^{N}\frac{1}{N_{\rm e}(\mathcal X_n)}\int_{\Omega_n}{\rm d}^3\mathbf r\left|\hat{\rho}(\mathbf r;\mathcal X_n)-\rho(\mathbf r;\mathcal X_n)\right|.
    \label{eq:nmae}
\end{align}

As shown in Tab.~\ref{tab:density-nmae}, AIDEN achieves SOTA performance under both exchange-correlation functionals on the general periodic-crystal benchmark, outperforming existing methods. On the QM9 molecular dataset, AIDEN ranks second, surpassed only by BOA, which is specifically designed for molecular systems. The lower accuracy on HSE than on PBE is expected. First, the HSE training set is substantially smaller, and the ECD benchmark likewise shows that charge-density prediction error decreases as the proportion of HSE data increases~\citep{chen2025ecd}. Second, PBE is a semilocal GGA functional whose exchange-correlation energy mainly depends on the local charge density and its gradient~\citep{pbe}, whereas HSE additionally incorporates screened Hartree-Fock exact exchange~\citep{heyd2003hse,krukau2006hse06}. This makes the density more sensitive to orbital localization and complex electronic environments, increasing the difficulty of learning the structure-to-HSE-density mapping.

\begin{figure}[t]
   \centering
   \includegraphics[width=\linewidth]{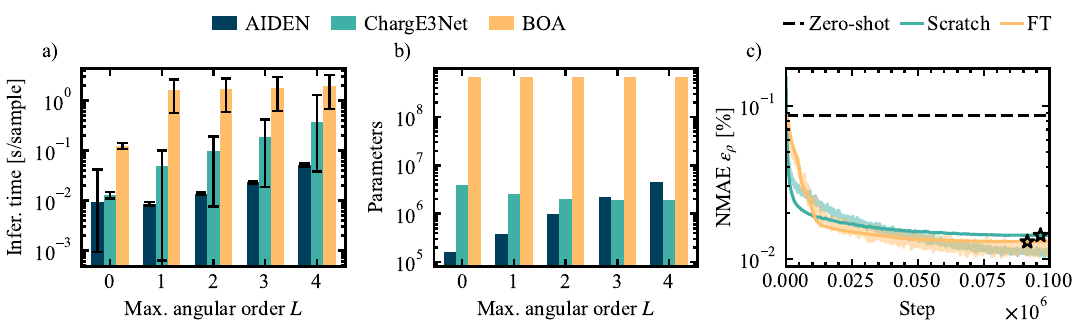}
   \caption{\textbf{Model efficiency, parameter count, and HSE transfer learning.} (a) Per-sample inference time of AIDEN, ChargE3Net, and BOA at different maximum angular orders $L$. Bar heights indicate the mean values, and the error bars denote one standard deviation. (b) Comparison of model parameter counts. (c) Comparison of AIDEN zero-shot error and two transfer-learning strategies. Opaque and semi-transparent curves denote validation and training losses, with pentagrams marking the best validation NMAE.}
   \label{fig:bar-hse}
   
\end{figure}

Fig.~\ref{fig:bar-hse}(a) shows that, at the same $L$, AIDEN provides substantially faster inference than ChargE3Net and BOA. This advantage further increases with system size as $N_{\rm a}$ grows (see Fig.~\ref{fig:tbg-all}(d)). Fig.~\ref{fig:bar-hse}(b) compares the dependence of parameter count on $L$. ChargE3Net actively reduces the number of channels assigned to each irreducible representation as $L$ increases~\citep{koker2024higher}, causing its total parameter count to decrease; however, whether this allocation is optimal for higher-order representation capacity remains unclear. In contrast, BOA maintains an almost constant parameter count by using a fixed feature-channel dimension while encoding angular dependence in non-parametric orbitals~\citep{boa}. As shown in Fig.~\ref{fig:bar-hse}(c), PBE pretraining yields a modest improvement, suggesting that the two functionals share the dominant density structure, while the remaining discrepancy is more closely related to HSE-specific exchange-correlation corrections. The lower performance of AIDEN than BOA on QM9 may partly result from the stronger task-specific priors used by BOA for molecular systems. BOA employs an element-specific \texttt{def2-QZVPPD} basis~\citep{basis} optimized for H/C/N/O/F, together with element-specific radial corrections, and performs message passing directly in the density basis through basis overlaps.

\begin{table}[t]
\centering
\caption{\textbf{Benchmarking charge density learning across different atomic systems and functionals.} All reported metrics are quantified by NMAE $\varepsilon_{\rho}$ (\%) (Eq.~\ref{eq:nmae}), with lower values indicating better model performance. The SOTA result is highlighted in bold, while the second-best result is underlined. The reference baselines include eqDeepDFT~\citep{jorgensen2022equivariant}, ChargE3Net~\citep{koker2024higher}, NeuralSCF~\citep{song2026neural}, SCDP~\citep{fu2024recipe}, ELECTRA~\citep{elsborg2025electra}, and BOA~\citep{boa}. For the PBE task, two $\varepsilon_{\rho}$ values are reported for both ChargE3Net and AIDEN, with the left and right entries corresponding to maximum angular order $L=3$ and $L=4$. For the HSE task, the two AIDEN results are obtained using the \textit{training-from-scratch} and \textit{fine-tuning} strategies.}
\label{tab:density-nmae}
\small
\begin{tabular}{lccccccc}
\toprule
Dataset & eqDeepDFT & ChargE3Net & NeuralSCF & SCDP & ELECTRA & BOA & \textbf{AIDEN} \\
\midrule
PBE & 0.799 & \underline{0.685} / \underline{0.523} & - & - & - & - & \textbf{0.621} / \textbf{0.4542} \\
HSE & - & \underline{1.534} & - & - & - & - & \textbf{1.4307 / 1.3012} \\
QM9 & 0.284 & 0.196 & 0.197 & 0.178 & 0.177 & \textbf{0.1339} & \underline{0.1628} \\
\bottomrule
\end{tabular}
\end{table}

\subsection{Generalization to OOD systems}

We evaluate zero-shot cross-system transfer of the PBE-pretrained model on these OOD systems, following the system choices of Ref.~\citep{eacnet}, without any system-specific fine-tuning. We compare charge densities and fixed-density energies and forces, and further evaluate non-self-consistent field (NSCF)~\citep{nscf-ref} bands for TBG.

\begin{figure}[t]
    \centering
    \includegraphics[width=.95\linewidth]{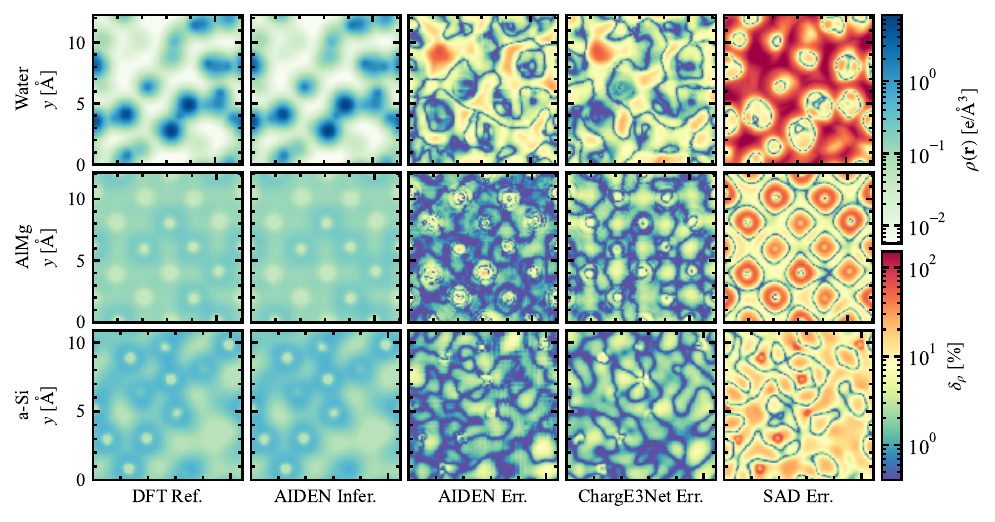}
    \caption{\textbf{Visualization of atom-rich charge density cross sections.} Rows correspond to Water, AlMg, and a-Si. The first two columns show the DFT reference and AIDEN-inferred $\rho(\mathbf{r})$; the remaining columns show the normalized pointwise deviation $\delta_{\rho}(\mathbf{r}_g):=|\hat{\rho}(\mathbf{r}_g)-\rho(\mathbf{r}_g)|/\rho(\mathbf{r}_g)$ of AIDEN, ChargE3Net, and SAD at grid point $g$. The density and deviation values are indicated by the upper and lower color bars, respectively. All atom-rich cross sections are perpendicular to the $x$ axis. Atoms within 1.5 grid spacings of each plane are counted, yielding H$_{\text{7}}$O$_{\text{2}}$, Al$_{\text{6}}$Mg$_{\text{7}}$, and Si$_{\text{5}}$ for Water, AlMg, and a-Si, respectively. Detailed settings are provided in Appx.~\ref{app:cross-sec}.}
    \label{fig:ood-density}
    
\end{figure}

\begin{table}[t]
\centering
\caption{\textbf{OOD benchmark.} Errors are relative to the matched PBE-SCF reference. The best and second-best errors are bold and underlined. Timing ranks compare the two models; SAD times ($\dagger$) include the complete fixed-density VASP calculation. Here, we define the per-atom energy error as $|\Delta E|:=|\hat{E}-E|/N_{\rm a}$, the mean absolute force component as $\langle |F_{i\alpha}| \rangle:=\frac{1}{3N_{\rm a}}\sum_{i=1}^{N_{\rm a}}\sum_{\alpha=\{x,y,z\}}|F_{i\alpha}|$, and the corresponding mean error as $\langle |\Delta F_{i\alpha}| \rangle:=\langle |\hat{F}_{i\alpha}-F_{i\alpha}| \rangle_{i,\alpha}$.}
\label{lab:ood-bench}
\setlength{\tabcolsep}{4pt}
\begin{tabular}{clcccc}
\toprule
System & Method & $\varepsilon_{\rho}$ [\%] $\downarrow$ & Time [s] $\downarrow$ & $|\Delta E|$ [meV/atom] $\downarrow$ & $\langle|\Delta F_{i\alpha}|\rangle$ [eV/\AA] $\downarrow$ \\
\midrule
\multirow{3}{*}{Water}
& AIDEN & \underline{1.8764} & \textbf{23.70} & \textbf{3.444} & \textbf{0.11467} \\
& ChargE3Net & \textbf{1.7880} & \underline{1286.43} & \underline{6.443} & \underline{0.12538} \\
& SAD & 12.3527 & $1433^{\dagger}$ & 485.445 & 1.11744 \\
\midrule
\multirow{3}{*}{AlMg}
& AIDEN & \textbf{0.9542} & \textbf{10.17} & \underline{3.484} & \underline{0.04977} \\
& ChargE3Net & \underline{1.1928} & \underline{486.82} & \textbf{0.938} & \textbf{0.02451} \\
& SAD & 10.2581 & $1382^{\dagger}$ & 53.269 & 0.14645 \\
\midrule
\multirow{3}{*}{a-Si}
& AIDEN & \textbf{1.3215} & \textbf{4.34} & \textbf{3.213} & \textbf{0.04087} \\
& ChargE3Net & \underline{1.6391} & \underline{328.87} & \underline{4.774} & \underline{0.04090} \\
& SAD & 9.9995 & $307^{\dagger}$ & 82.107 & 0.12778 \\
\bottomrule
\end{tabular}
\end{table}

\textbf{Water, AlMg and a-Si.}
For OOD evaluation, we consider a water cell containing 64 molecules, an Al$_\text{54}$Mg$_\text{54}$ alloy, and a 64-atom a-Si structure. Reference densities and properties are recomputed using PBE-SCF with a 520~eV cutoff; candidate densities are evaluated on the complete grids and held fixed for property calculations (Appx.~\ref{app:property}). As shown in Fig.~\ref{fig:ood-density} and Tab.~\ref{lab:ood-bench}, AIDEN gives the lowest full-grid $\varepsilon_{\rho}$ on AlMg and a-Si, while ChargE3Net performs slightly better on water. Both learned models substantially outperform SAD in density accuracy across all three systems. AIDEN evaluates the full grids in 4.34-23.70~s, corresponding to a measured 48-76$\times$ speedup over the ChargE3Net implementation used here. The fixed-density calculations further show that the predicted densities retain useful downstream electronic-structure information: AIDEN gives lower energy errors on water and a-Si and lower or nearly identical force errors on these two systems, while ChargE3Net performs better on AlMg. Interestingly, a lower $\varepsilon_{\rho}$ does not necessarily imply smaller $|\Delta E|$ or $\langle|\Delta F_{i\alpha}|\rangle$, since NMAE discards the sign and spatial distribution of the density error, whereas downstream observables weight different regions differently and can exhibit error cancellation~\citep{li2025image,10.1021/acs.jctc.7b00550}. We therefore treat energies and forces as complementary measures of density fidelity; further discussion is provided in Appx.~\ref{app:nmae-downstream}.


\begin{figure}[t]
   \centering
   \includegraphics[width=\linewidth]{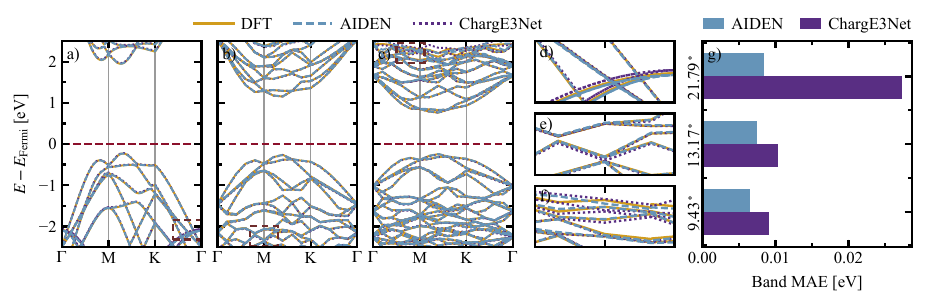}
   \caption{\textbf{NSCF band structures of TBG.} (a) $21.79^\circ$, $N_{\rm a}=28$; (b) $13.17^\circ$, $N_{\rm a}=76$; and (c) $9.43^\circ$, $N_{\rm a}=148$. Yellow solid, blue dashed, and purple dashed curves denote DFT, AIDEN, and ChargE3Net. (d), (e), and (f) show the regions with pronounced errors produced by ChargE3Net in the three TBG band structures, corresponding to the dashed boxes in (a)-(c). (g) Error comparison of TBG band structures computed from the charge densities predicted by AIDEN and ChargE3Net against the DFT reference.}
   \label{fig:tbg-band}
   
\end{figure}

\textbf{NSCF band structure calculation of TBG.}
We further evaluate TBG at $21.79^\circ$, $13.17^\circ$, and $9.43^\circ$. Fixed-density PBE calculations use \texttt{ICHARG=11}, a 520~eV cutoff, and the $\Gamma$--M--K--$\Gamma$ path. Fig.~\ref{fig:tbg-band} shows bands with the DFT $E_{\rm Fermi}$ set to zero and one least-squares rigid shift applied per model and cell. Within $E_{\rm Fermi}\pm5$~eV, the aligned band MAEs are 8.42, 7.41, and 11.47~meV for AIDEN, versus 27.35, 10.30, and 12.93~meV for ChargE3Net. These aligned MAEs quantify residual band-shape and dispersion errors after removal of a global reference-energy offset. Full-grid inference takes 13.27-41.43~s, a measured 9.7-15.3$\times$ speedup over the evaluated ChargE3Net implementation. Appx.~\ref{app:tbg-results} reports the shifts, raw and tail-error statistics, sampling differences, and timing protocol.

\section{Discussion}

In this work, we introduced AIDEN for accurate and scalable real-space charge density learning. At its core, AIDEN adopts a physically inspired, learnable decomposition that separates an element-dependent one-center baseline from environment-induced charge redistribution. The former captures the dominant local density structure shared across different environments, while the latter describes bonding- and coordination-dependent corrections, providing a natural inductive bias across diverse chemical and structural spaces. This decomposition is integrated with an equivariant encoder and a continuous decoder to preserve the required geometric symmetries while enabling efficient field evaluation. Experiments demonstrate high accuracy on periodic-crystal and molecular benchmarks, effective PBE-to-HSE fine-tuning, and zero-shot transfer to structurally distinct OOD systems. Fixed-density calculations further assess the predicted densities through energies and forces, as well as band structures in TBG. Full-grid inference shows favorable scalability to millions of grid points. Notably, TBG electron counts deviate by only 0.0068-0.0095\% without explicit normalization (Appx.~\ref{app:charge-count}).

\textbf{Future perspective.} AIDEN currently focuses on GS scalar charge densities. Future work may extend it to spin-resolved densities and broader electronic-structure settings, including unified models across exchange-correlation functionals and chemical spaces. Joint learning of charge densities with energies, forces, and other DFT observables may further improve electronic-structure fidelity and generalization.

\section{Code availability}


The code supporting this work is available in the \href{https://github.com/FlorianTseng/AIDEN}{AIDEN repository}. The PBE and HSE datasets used for model training are provided by \href{https://github.com/pincher-chen/ECDBench}{ECDBench}, the QM9 dataset is available from the \href{https://data.dtu.dk/articles/dataset/QM9_Charge_Densities_and_Energies_Calculated_with_VASP/16794500}{QM9 charge density dataset}.

\section{Acknowledgements}

We would like to acknowledge the National Key R\&D Program of China (No. 2021YFA0718900), the National Natural Science Foundation of China (Nos. 12374096 and 92477114), and the Jiangsu Funding Program for Excellent Postdoctoral Talent for financial support. Z. Zhong thanks the Suzhou Innovation and Entrepreneurship Leading Talent Program and the Gusu Leadership Program for their support.

\section{The use of large language models}

In this work, we used large language models to assist with language polishing and manuscript editing. We have reviewed all AI-assisted content and take full responsibility for the final manuscript.

\bibliography{ref}
\bibliographystyle{iclr2027_conference}

\newpage
\appendix

{\huge Appendix}

\section{Supplementary theory}

\subsection{Kohn-Sham density functional theory and charge density self-consistency}
\label{app:ks-scf}

In this section, we summarize the electronic-structure relations underlying the reference electron densities used throughout this work~\citep{RevModPhys.87.897,scf-ref}. For a fixed atomic structure $\mathcal{X}$ within the Born-Oppenheimer approximation, the interacting electronic problem is governed, up to the ion-ion contribution, by the many-electron Hamiltonian
\begin{align}
\hat{H}_{\mathcal{X}}=\sum_{a=1}^{N_{\mathrm{e}}}\left[-\frac{1}{2}\nabla_a^2+v_{\mathrm{ext}}^{\mathcal{X}}(\mathbf{r}_a)\right]+\frac{1}{2}\sum_{a\ne b}\frac{1}{|\mathbf{r}_a-\mathbf{r}_b|},
\label{eq:many-electron-hamiltonian}
\end{align}
where $N_{\mathrm{e}}$ is the number of electrons, $\mathbf{r}_a$ is the coordinate of electron $a$, $v_{\mathrm{ext}}^{\mathcal{X}}$ is the external potential generated by the fixed ionic configuration $\mathcal{X}$, atomic units are used, and the standard periodic electrostatic convention is understood for crystalline systems. Suppressing spin variables for notational clarity, an interacting GS wavefunction $\Psi_{\mathcal{X}}^{\star}$ determines the one-particle reduced density matrix and its diagonal electron density as
\begin{align}
\gamma_{\mathcal{X}}(\mathbf{r},\mathbf{r}')&=N_{\mathrm{e}}\int\prod_{i=2}^{N_{\mathrm{e}}}d\mathbf{r}_i\Psi_{\mathcal{X}}^{\star}(\mathbf{r},\mathbf{r}_2,\ldots,\mathbf{r}_{N_{\mathrm{e}}})\overline{\Psi_{\mathcal{X}}^{\star}(\mathbf{r}',\mathbf{r}_2,\ldots,\mathbf{r}_{N_{\mathrm{e}}})},\\
\rho_{\mathcal{X}}^{\star}(\mathbf{r})&=\gamma_{\mathcal{X}}(\mathbf{r},\mathbf{r}),\quad \int_{\Omega}\rho_{\mathcal{X}}^{\star}(\mathbf{r})d^3\mathbf{r}=N_{\mathrm{e}}.
\label{eq:one-rdm-density}
\end{align}
Here, $\gamma_{\mathcal{X}}(\mathbf{r},\mathbf{r}')$ is the one-particle reduced density matrix and $\rho_{\mathcal{X}}^{\star}(\mathbf{r})$ is the corresponding GS electron density over the unit cell $\Omega$. Thus, the real-space density is the diagonal of a more general nonlocal one-particle object. Within the Hohenberg-Kohn framework, the Levy constrained-search formulation removes the many-electron wavefunction from the explicit GS variational variable by defining the universal functional $F_{\mathrm{HK}}[\rho]=\min_{\Psi\to\rho}\langle\Psi|\hat{T}+\hat{W}_{\mathrm{ee}}|\Psi\rangle$. The electronic GS energy is then obtained as
\begin{align}
E_0=\min_{\rho\in\mathcal{D}_{N_{\mathrm{e}}}}\left\{F_{\mathrm{HK}}[\rho]+\int_{\Omega}v_{\mathrm{ext}}^{\mathcal{X}}(\mathbf{r})\rho(\mathbf{r})d^3\mathbf{r}\right\}.
\label{eq:hk-variational}
\end{align}
Here, $\hat{T}$ and $\hat{W}_{\mathrm{ee}}$ denote the many-electron kinetic-energy and electron-electron interaction operators, respectively, $\Psi\to\rho$ indicates wavefunctions yielding the density $\rho$, and $\mathcal{D}_{N_{\mathrm{e}}}$ denotes the set of admissible densities integrating to $N_{\mathrm{e}}$ electrons. This variational statement establishes the GS electron density as a sufficient fundamental variable for the electronic GS problem, even though the underlying interacting state is described by a many-electron wavefunction.

KS-DFT makes the density functional computationally tractable by introducing an auxiliary non-interacting system with the same GS density. Up to the ion-ion contribution, its total energy is written as
\begin{align}
E_{\mathcal{X}}[\rho]=T_{\mathrm{s}}[\rho]+E_{\mathrm{ext}}^{\mathcal{X}}[\rho]+E_{\mathrm{H}}[\rho]+E_{\mathrm{xc}}[\rho],
\label{eq:ks-energy-functional}
\end{align}
where $T_{\mathrm{s}}$ is the non-interacting kinetic energy, $E_{\mathrm{ext}}^{\mathcal{X}}$ is the interaction with the external ionic potential, $E_{\mathrm{H}}$ is the classical electron-electron electrostatic energy, and $E_{\mathrm{xc}}$ contains the remaining exchange and correlation effects. Writing $v_{\mathrm{C}}^{\mathrm{per}}$ for the periodic Coulomb kernel, the Hartree term and its functional derivative are
\begin{align}
E_{\mathrm{H}}[\rho]=\frac{1}{2}\iint_{\Omega\times\Omega}\rho(\mathbf{r})v_{\mathrm{C}}^{\mathrm{per}}(\mathbf{r}-\mathbf{r}')\rho(\mathbf{r}')d^3\mathbf{r}d^3\mathbf{r}',\quad v_{\mathrm{H}}[\rho](\mathbf{r})=\int_{\Omega}v_{\mathrm{C}}^{\mathrm{per}}(\mathbf{r}-\mathbf{r}')\rho(\mathbf{r}')d^3\mathbf{r}'.
\label{eq:hartree-functional}
\end{align}
The Hartree potential therefore couples the density at $\mathbf{r}$ to the density throughout the cell. Taking the functional derivative of Eq.~\ref{eq:ks-energy-functional} gives the density-dependent KS Hamiltonian
\begin{align}
H_{\mathrm{KS}}[\rho]=-\frac{1}{2}\nabla^2+v_{\mathrm{ext}}^{\mathcal{X}}(\mathbf{r})+v_{\mathrm{H}}[\rho](\mathbf{r})+v_{\mathrm{xc}}[\rho](\mathbf{r}),\quad v_{\mathrm{xc}}[\rho](\mathbf{r})=\frac{\delta E_{\mathrm{xc}}}{\delta\rho(\mathbf{r})}.
\label{eq:ks-hamiltonian}
\end{align}
For a periodic system, solving the KS equations at each sampled $\mathbf{k}$ point gives
\begin{align}
H_{\mathrm{KS}}[\rho]\phi_{n\mathbf{k}}(\mathbf{r})=\epsilon_{n\mathbf{k}}\phi_{n\mathbf{k}}(\mathbf{r}),\quad \rho(\mathbf{r})=\sum_{\mathbf{k}}\sum_n w_{\mathbf{k}}f_{n\mathbf{k}}|\phi_{n\mathbf{k}}(\mathbf{r})|^2,\quad \sum_{\mathbf{k}}\sum_n w_{\mathbf{k}}f_{n\mathbf{k}}=N_{\mathrm{e}},
\label{eq:ks-orbital-density}
\end{align}
where $\phi_{n\mathbf{k}}$ and $\epsilon_{n\mathbf{k}}$ are the KS Bloch orbital and its eigenvalue for band $n$ and wavevector $\mathbf{k}$, while $w_{\mathbf{k}}$ and $f_{n\mathbf{k}}$ denote the Brillouin-zone weight and orbital occupation, respectively. The Hamiltonian is linear for a fixed input density, but the complete KS problem is nonlinear because $v_{\mathrm{H}}$ and $v_{\mathrm{xc}}$ depend on the density reconstructed from its eigenstates.

The distinction between a local density and a nonlocal one-particle density matrix becomes particularly clear when comparing semilocal and hybrid functionals. PBE uses a generalized-gradient form $E_{\mathrm{xc}}^{\mathrm{PBE}}[\rho,\nabla\rho]$~\citep{pbe}, whereas Hartree-Fock exchange acts nonlocally on an orbital through the occupied-state density matrix. With $\gamma_{\mathrm{s}}(\mathbf{r},\mathbf{r}')=\sum_{\mathbf{k}n}w_{\mathbf{k}}f_{n\mathbf{k}}\phi_{n\mathbf{k}}(\mathbf{r})\phi_{n\mathbf{k}}^{*}(\mathbf{r}')$ and spin labels suppressed, a screened Fock exchange operator may be written as
\begin{align}
\left(\hat{V}_{x}^{\mathrm{HF,SR}}[\gamma_{\mathrm{s}}]\psi\right)(\mathbf{r})=-\int_{\Omega}\gamma_{\mathrm{s}}(\mathbf{r},\mathbf{r}')w_{\omega}^{\mathrm{SR}}(|\mathbf{r}-\mathbf{r}'|)\psi(\mathbf{r}')d^3\mathbf{r}',\quad w_{\omega}^{\mathrm{SR}}(r)=\frac{\mathrm{erfc}(\omega r)}{r}.
\label{eq:hf-screened-exchange}
\end{align}
Here, $\gamma_{\mathrm{s}}$ is the one-particle density matrix of the auxiliary non-interacting system, $\psi$ denotes the orbital on which the nonlocal operator acts, $\omega$ is the range-separation parameter, and $\mathrm{SR}$ denotes the short-range part of the Coulomb interaction. HSE combines this nonlocal short-range exchange with semilocal PBE exchange and correlation~\citep{heyd2003hse,krukau2006hse06},
\begin{align}
E_{\mathrm{xc}}^{\mathrm{HSE}}=\alpha E_{x}^{\mathrm{HF,SR}}(\omega)+(1-\alpha)E_{x}^{\mathrm{PBE,SR}}(\omega)+E_{x}^{\mathrm{PBE,LR}}(\omega)+E_{c}^{\mathrm{PBE}},
\label{eq:hse-xc}
\end{align}
where $\alpha$ is the screened Hartree-Fock exchange fraction and $\mathrm{LR}$ denotes the complementary long-range contribution. HSE is therefore most naturally formulated as a generalized KS problem whose effective operator depends on both the density and the occupied-state one-particle density matrix,
\begin{align}
H_{\mathrm{gKS}}^{\mathrm{HSE}}[\rho,\gamma_{\mathrm{s}}]=-\frac{1}{2}\nabla^2+v_{\mathrm{ext}}^{\mathcal{X}}+v_{\mathrm{H}}[\rho]+v_{\mathrm{xc}}^{\mathrm{HSE,sl}}[\rho]+\alpha\hat{V}_{x}^{\mathrm{HF,SR}}[\gamma_{\mathrm{s}}],
\label{eq:hse-gks-hamiltonian}
\end{align}
where $v_{\mathrm{xc}}^{\mathrm{HSE,sl}}$ denotes the remaining semilocal HSE exchange-correlation potential associated with $(1-\alpha)E_{x}^{\mathrm{PBE,SR}}+E_{x}^{\mathrm{PBE,LR}}+E_{c}^{\mathrm{PBE}}$. This distinction is relevant to the PBE and HSE datasets considered in this work: the PBE effective potential is determined by semilocal density information, whereas HSE additionally depends on the occupied orbital manifold through nonlocal screened exchange. In both cases, however, the converged observable learned by AIDEN remains the same basis-independent real-space scalar field $\rho_{\mathcal{X}}^{\star}(\mathbf{r})$.

The SCF construction can also be written directly in matrix form. Expanding the orbitals in an arbitrary finite basis $\boldsymbol{\chi}_{\mathbf{k}}(\mathbf{r})$ gives the generalized eigenvalue problem
\begin{align}
\mathbf{H}[\mathbf{D}](\mathbf{k})\mathbf{C}(\mathbf{k})=\mathbf{S}(\mathbf{k})\mathbf{C}(\mathbf{k})\boldsymbol{\epsilon}(\mathbf{k}),\quad \mathbf{C}^{\dagger}(\mathbf{k})\mathbf{S}(\mathbf{k})\mathbf{C}(\mathbf{k})=\mathbf{I},
\label{eq:ks-generalized-eigenproblem}
\end{align}
where $\boldsymbol{\chi}_{\mathbf{k}}(\mathbf{r})$ collects the basis functions at wavevector $\mathbf{k}$, $\mathbf{H}[\mathbf{D}](\mathbf{k})$ is the Hamiltonian matrix in this basis, $\mathbf{C}(\mathbf{k})$ contains the corresponding orbital coefficients, $\boldsymbol{\epsilon}(\mathbf{k})$ is the diagonal matrix of orbital eigenvalues, and $\mathbf{S}(\mathbf{k})$ is the basis-overlap matrix, which reduces to the identity for an orthonormal basis. The occupied eigenvectors define the one-particle density matrix and the corresponding real-space density,
\begin{align}
\mathbf{D}(\mathbf{k})=\mathbf{C}(\mathbf{k})\mathbf{f}(\mathbf{k})\mathbf{C}^{\dagger}(\mathbf{k}),\quad \rho(\mathbf{r})=\sum_{\mathbf{k}}w_{\mathbf{k}}\boldsymbol{\chi}_{\mathbf{k}}^{\dagger}(\mathbf{r})\mathbf{D}(\mathbf{k})\boldsymbol{\chi}_{\mathbf{k}}(\mathbf{r}),
\label{eq:density-matrix-to-density}
\end{align}
where $\mathbf{f}(\mathbf{k})$ is the diagonal occupation matrix and $\mathbf{D}(\mathbf{k})$ is the one-particle density matrix in the chosen basis. For semilocal KS-DFT, $\mathbf{H}$ depends on $\mathbf{D}$ through the reconstructed $\rho$; for hybrid generalized KS calculations, the nonlocal exchange term additionally depends on off-diagonal information in $\mathbf{D}$ itself. The numerical matrix representation of $\mathbf{D}$ and $\mathbf{H}$ changes with the orbital basis, whereas the reconstructed scalar field $\rho(\mathbf{r})$ does not. This distinction is one of the motivations for directly learning the real-space charge density in AIDEN.

Although $\rho_{\mathcal{X}}^{\star}$ is defined variationally, practical calculations normally obtain it through a fixed-point SCF iteration. Starting from an input density $\rho_{\mathcal{X}}^{(t)}$, a semilocal KS calculation constructs $H_{\mathrm{KS}}[\rho_{\mathcal{X}}^{(t)}]$, solves the eigenproblem, and reconstructs an output density $\widetilde{\rho}_{\mathcal{X}}^{(t+1)}$. A simple density-mixing step is
\begin{align}
\widetilde{\rho}_{\mathcal{X}}^{(t+1)}=\mathcal{K}_{\mathcal{X}}\left[\rho_{\mathcal{X}}^{(t)}\right],\quad \rho_{\mathcal{X}}^{(t+1)}=(1-\alpha_{\mathrm{mix}})\rho_{\mathcal{X}}^{(t)}+\alpha_{\mathrm{mix}}\widetilde{\rho}_{\mathcal{X}}^{(t+1)},
\label{eq:scf-density-update}
\end{align}
where $\mathcal{K}_{\mathcal{X}}$ denotes one KS density-update map, $\widetilde{\rho}_{\mathcal{X}}^{(t+1)}$ is the unmixed output density, and $\alpha_{\mathrm{mix}}$ is the density-mixing parameter. Pulay- and quasi-Newton-type schemes replace this simple mixing operation in practical codes, while hybrid calculations additionally update the occupied-state density matrix entering the Fock term. In either case, convergence is a fixed-point condition, $\rho_{\mathcal{X}}^{\star}=\mathcal{K}_{\mathcal{X}}[\rho_{\mathcal{X}}^{\star}]$, together with the corresponding consistency of the occupied orbitals or density matrix. Following Eq.~\ref{eq:scf}, the complete relation can be summarized as
\begin{align}
(\rho_{\mathcal{X}}^{(t)},\mathbf{D}^{(t)})\rightarrow H_{\mathrm{eff}}[\rho_{\mathcal{X}}^{(t)},\mathbf{D}^{(t)}]\rightarrow\{\phi_{n\mathbf{k}}^{(t+1)},\epsilon_{n\mathbf{k}}^{(t+1)}\}\rightarrow\mathbf{D}^{(t+1)}\rightarrow\widetilde{\rho}_{\mathcal{X}}^{(t+1)}(\mathbf{r})\rightarrow\rho_{\mathcal{X}}^{(t+1)}(\mathbf{r}).
\label{eq:scf-cycle-appendix}
\end{align}
Here, $H_{\mathrm{eff}}$ denotes the effective single-particle operator, corresponding to the KS Hamiltonian for a semilocal functional or the generalized KS operator for a hybrid functional. For PBE, the explicit $\mathbf{D}$ dependence of $H_{\mathrm{eff}}$ reduces to its dependence through $\rho$, recovering the density-only form in Eq.~\ref{eq:scf}; for HSE, the off-diagonal one-particle information also enters through screened exact exchange. The charge density therefore plays two roles simultaneously: it determines a central part of the effective electronic Hamiltonian and is reconstructed from the eigenstates of that same Hamiltonian. This mutual dependence is the origin of charge-density self-consistency and motivates learning $\rho_{\mathcal{X}}^{\star}$ as a direct electronic-structure target.


\subsection{Density target and fixed-density evaluation}
\label{app:density-target}

The formal density in Eq.~\ref{eq:physical-density} motivates learning a real-space field. In the numerical calculations, AIDEN predicts the FFT-grid density stored by VASP. Its integral gives the corresponding electron count.

For the model-based OOD property calculations, the predicted grid density is used without charge renormalization in fixed-density VASP calculations. The remaining restart quantities and electronic settings are kept identical between AIDEN and ChargE3Net within each structure, while SAD uses its own atomic initialization. With \texttt{ICHARG=11}, VASP holds the supplied density fixed during electronic minimization~\citep{vasp-icharg}. The reported energies, forces, and spectra are obtained from these fixed-density calculations. The forces are the fixed-density estimates reported by VASP. 

\subsection{Theory of superposition of atomic densities}
\label{app:sad}

We briefly introduce the physical motivation of the SAD, which is closely related to the density decomposition adopted in AIDEN. The central approximation of SAD is to regard a many-atom system, at zeroth order, as a collection of isolated atoms and construct its initial electron density by adding the densities of the constituent atoms~\citep{sad}. For an isolated atom of species $Z$, let $\rho_Z^{\rm atom}(\mathbf r)=\rho_Z^{\rm atom}(|\mathbf r|)$ denote its spherically averaged atomic density. For the periodic structure $\mathcal X=(\mathbf A,\{Z_i,\mathbf s_i\}_{i=1}^{N_{\rm a}})$, the corresponding SAD density can be written as
\begin{align}
\rho_{\rm SAD}(\mathbf r;\mathcal X)=\sum_{i=1}^{N_{\rm a}}\sum_{\mathbf n\in\mathbb Z^3}
\rho_{Z_i}^{\rm atom}\left[\mathbf r-(\mathbf R_i+\mathbf n\mathbf A)\right].
\label{eq:sad-density}
\end{align}
Each contribution depends only on the atomic species and the displacement from its center, so $\rho_{\rm SAD}$ contains no explicit information about the surrounding chemical environment.

The physical basis of this approximation is that a substantial fraction of the real-space density is already determined by the element-dependent atomic electronic structure. In particular, the strong density variation around each nucleus is largely inherited from the corresponding atomic shells. SAD therefore provides a physically meaningful one-center baseline before bonding and other interatomic effects are taken into account. This idea has long been used to initialize SCF calculations, where atomic densities are first generated separately and then combined to construct the molecular density matrix~\citep{sad}. A closely related independent-atom approximation can also be made at the potential level: in the superposition of atomic potentials (SAP), the effective one-particle potential of the full system is approximated by a sum of atomic effective potentials~\citep{lehtola2020}. SAD and SAP operate on different physical quantities, but both rely on the same zeroth-order picture that the dominant local electronic structure can be inherited from isolated atoms.

\begin{figure}[t]
    \centering
    \includegraphics[width=.7\linewidth]{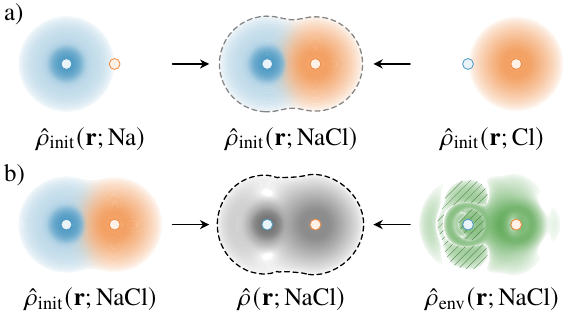}
    \caption{\textbf{Illustration of the components of the AIDEN-predicted charge density $\hat{\rho}(\mathbf{r})$.} (a) The $\hat{\rho}_{\rm init}(\mathbf{r}; Z)$ branch, whose shape depends only on the elemental species $Z$, is first initialized through pretraining and subsequently optimized jointly with $\hat{\rho}_{\rm env}(\mathbf{r})$. (b) The $\hat{\rho}_{\rm env}(\mathbf{r})$ branch, which is responsible for expressing complex spatial charge-density patterns and can also be interpreted as a refinement of $\hat{\rho}_{\rm init}(\mathbf{r}; Z)$; accordingly, $\hat{\rho}_{\rm env}(\mathbf{r})$ may take signed values and is shown here with dashed lines. We use a fictitious NaCl diatomic unit cell for illustration. Blue, orange, and green denote Na, Cl, and the charge density contributed by the local-environment branch, respectively. The displayed values are actual inferences from a pretrained AIDEN model with $L=4$, and the final total charge density satisfies $\hat{\rho}(\mathbf{r};\text{NaCl})>0$.}
    \label{fig:sad-illu}
\end{figure}

The SAD density is not, however, the self-consistent GS density of the interacting system. Once the atoms are brought together, the KS potential depends on the complete environment and the density is redistributed through chemical bonding, hybridization, polarization, screening, and charge transfer. We can therefore write
\begin{align}
\rho_{\mathcal X}^{\star}(\mathbf r)=\rho_{\rm SAD}(\mathbf r;\mathcal X)+\Delta\rho_{\mathcal X}(\mathbf r),
\label{eq:sad-redistribution}
\end{align}
where $\Delta\rho_{\mathcal X}$ denotes the environment-induced redistribution relative to the independent-atom reference. When the atomic reference and the self-consistent density contain the same number of electrons, this redistribution only rearranges charge in real space,
\begin{align}
\int_{\Omega}\Delta\rho_{\mathcal X}(\mathbf r)\,{\rm d}^3\mathbf r=0.
\label{eq:sad-charge-conservation}
\end{align}
Accordingly, SAD should be interpreted as an initial physical reference rather than an approximation to the fully converged density itself. In the original SAD construction, the summed atomic density matrix is generally non-idempotent and is used to construct a starting Fock or KS operator before the usual SCF iterations recover self-consistency~\citep{sad}. This picture directly motivates the decomposition used in AIDEN. In our case, we do not hard-code a fixed SAD density. Instead, we represent the predicted density as
\begin{align}
\widehat\rho_{\boldsymbol\theta}(\mathbf r;\mathcal X)=\rho_{\rm init}(\mathbf r)+\rho_{\rm env}(\mathbf r),
\label{eq:sad-aiden}
\end{align}
where $\rho_{\rm init}$ is a learnable element-dependent one-center contribution and $\rho_{\rm env}$ introduces the environment-dependent correction. The former inherits the independent-atom intuition of SAD, while the latter accounts for the density redistribution that cannot be described by isolated atomic densities. Since $\rho_{\rm init}$ is learned jointly with the complete model after a short pretraining stage, it should not be identified with a fixed SAD reference. Rather, AIDEN generalizes the SAD picture into a learnable decomposition in which both the atomic baseline and the environment-induced contribution are optimized directly from self-consistent reference densities.

\subsection{Spherical-harmonic representation}
\label{app:spherical-harmonics}

Spherical harmonics provide a natural angular basis for representing 3D geometric information under rotations. For angular order $\ell$, the corresponding irreducible representation of SO(3) contains $2\ell+1$ independent components indexed by $m=-\ell,\dots,\ell$. In AIDEN, however, we do not explicitly store the conventional complex spherical harmonics $Y_{\ell m}$. Instead, we use their equivalent irreducible Cartesian representation throughout the Cartesian encoder. Given a unit direction $\widehat{\mathbf d}$, the rank-$\ell$ Cartesian harmonic used in our implementation is
\begin{align}
\mathbf Y^{(\ell)}(\widehat{\mathbf d})=\frac{(2\ell-1)!!}{\ell!}\mathcal P_{\ell}^{\rm irr}\left[\widehat{\mathbf d}^{\otimes\ell}\right],\quad \mathbf Y^{(0)}=1,
\label{eq:cartesian-harmonic}
\end{align}
where $\mathcal P_{\ell}^{\rm irr}$ projects the tensor product onto the fully symmetric traceless rank-$\ell$ subspace. Thus, although $\mathbf Y^{(\ell)}$ is stored using $3^\ell$ Cartesian entries, permutation symmetry and the traceless constraints leave exactly $2\ell+1$ independent degrees of freedom. Under a rotation $\mathbf R\in{\rm SO}(3)$, it transforms as $\mathbf Y^{(\ell)}(\mathbf R\widehat{\mathbf d})=\mathbf R^{\otimes\ell}\mathbf Y^{(\ell)}(\widehat{\mathbf d})$, while inversion gives $\mathbf Y^{(\ell)}(-\widehat{\mathbf d})=(-1)^\ell\mathbf Y^{(\ell)}(\widehat{\mathbf d})$. Thus the angular dependence and its natural parity are fixed analytically, and no individual Cartesian component needs to be learned.

\begin{figure}[t]
    \centering
    \includegraphics[width=.75\linewidth]{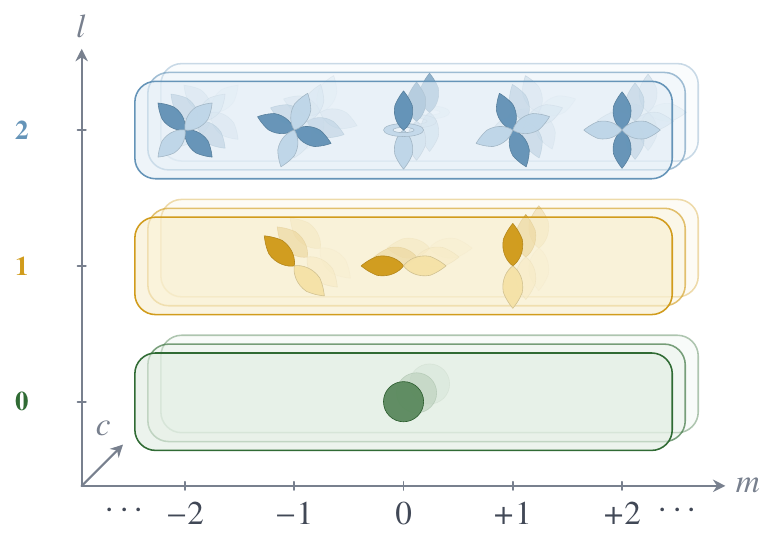}
    \caption{Illustration of the real spherical-harmonic basis for angular orders $\ell=0,1,2$. Each order contains $2\ell+1$ independent components indexed by $m=-\ell,\ldots,\ell$, corresponding to the irreducible SO(3) subspace represented by the symmetric-traceless Cartesian tensors in AIDEN. The two colors indicate the positive and negative signs of the angular basis functions.}
    \label{fig:sph-demo}
\end{figure}

The same irreducible tensor can alternatively be expressed in a compact real-spherical basis. Let $\mathcal U_{\ell}$ denote the fixed orthogonal change of basis used by the model. We then have
\begin{align}
\mathbf y^{(\ell)}=\mathcal U_{\ell}{\rm vec}\left[\mathbf Y^{(\ell)}\right]\in\mathbb R^{2\ell+1},\quad \mathbf Y^{(\ell)}={\rm vec}^{-1}\left[\mathcal U_{\ell}^{\dagger}\mathbf y^{(\ell)}\right],\quad \mathcal U=\bigoplus_{\ell=0}^{L}\mathcal U_{\ell}.
\label{eq:cartesian-spherical-map}
\end{align}
So, the Cartesian and real-spherical forms do not represent different physical information; they are simply two bases of the same angular irreducible space. The Cartesian form is convenient for the tensor products, contractions, and symmetric-traceless projections used in GIE and Cartesian ACE, whereas the compact spherical form is convenient for the edge-aligned SO(2) operations in TECE and for density decoding. The transformation $\mathcal U$ is fixed by the angular algebra and contains no learnable parameters.

The resulting hierarchy is intuitive. For $\ell=0$, there is only one isotropic scalar component; $\ell=1$ contains three vector-like components; and $\ell=2$ contains five quadrupolar components, corresponding to the familiar $s$-, $p$-, and $d$-like angular patterns illustrated in our schematic. More generally, each angular order $\ell$ contains $2\ell+1$ components, while the channel index $c=1,\dots,C$ simply provides multiple learned copies of the same representation and does not change its rotational transformation law. The positive and negative lobes in the schematic indicate the sign of the real angular basis functions rather than different atomic densities or physical orbitals. In this way, learned channel amplitudes encode the chemical environment, while the analytic harmonic basis supplies the required angular structure, SO(3) transformation law, and natural parity $(-1)^\ell$.

\subsection{Equivariance of AIDEN}
\label{app:equivariance}

We characterize the geometric transformation law of AIDEN under Euclidean motions. Let $g=(\mathbf{Q},\mathbf{t})$ denote a rigid transformation, where $\mathbf{Q}\in\mathrm{O}(3)$ is an orthogonal matrix satisfying $\mathbf{Q}^{\top}\mathbf{Q}=\mathbf{I}$ and $\mathbf{t}\in\mathbb{R}^{3}$ is a translation vector. We use row-vector coordinates throughout, so a spatial point transforms as $g\mathbf{r}=\mathbf{r}\mathbf{Q}^{\top}+\mathbf{t}$. Proper rotations satisfy $\det(\mathbf{Q})=+1$ and form $\mathrm{SO}(3)$, while proper rotations together with translations form $\mathrm{SE}(3)$. For a periodic structure $\mathcal{X}$ with lattice matrix $\mathbf{A}$ and atomic position $\mathbf{R}_i$, the transformed quantities are $\mathbf{A}'=\mathbf{A}\mathbf{Q}^{\top}$ and $\mathbf{R}_i'=\mathbf{R}_i\mathbf{Q}^{\top}+\mathbf{t}$. Therefore a periodic image $\mathbf{R}_i+\mathbf{n}\mathbf{A}$, with lattice-image index $\mathbf{n}\in\mathbb{Z}^{3}$, transforms as $(\mathbf{R}_i+\mathbf{n}\mathbf{A})\mathbf{Q}^{\top}+\mathbf{t}$. All geometric inputs to AIDEN are constructed from relative displacements, so the translation $\mathbf{t}$ cancels exactly. We first derive the transformation law for $\mathbf{Q}\in\mathrm{SO}(3)$ and then verify spatial inversion numerically. This covers the complete numerical $\mathrm{E}(3)$ behavior because every improper matrix $\mathbf{Q}\in\mathrm{O}(3)$ with $\det(\mathbf{Q})=-1$ can be written as $\mathbf{Q}=(-\mathbf{I})\mathbf{R}$ with $\mathbf{R}\in\mathrm{SO}(3)$.

\textbf{Irreducible representation matrices and Cartesian encoder.} For a rank-$\ell$ Cartesian tensor $\mathbf{T}$, let $\mathscr{D}_{\ell}(\mathbf{Q})$ denote the action of $\mathbf{Q}$ on its Cartesian indices,
\begin{align}
\left[\mathscr{D}_{\ell}(\mathbf{Q})\mathbf{T}\right]_{a_1\ldots a_{\ell}}=\sum_{b_1,\ldots,b_{\ell}}Q_{a_1b_1}\cdots Q_{a_{\ell}b_{\ell}}T_{b_1\ldots b_{\ell}},\quad \mathscr{D}_{0}(\mathbf{Q})=1.
\label{eq:aiden-cartesian-action}
\end{align}
Here, $\ell$ is the angular order and $a_1,\ldots,a_{\ell}$ and $b_1,\ldots,b_{\ell}$ label Cartesian components. Restricting $\mathscr{D}_{\ell}$ to the symmetric traceless rank-$\ell$ subspace gives the $(2\ell+1)$-dimensional irreducible representation of $\mathrm{SO}(3)$. We denote its matrix in an orthonormal irreducible basis by $\mathbf{D}^{(\ell)}(\mathbf{Q})\in\mathbb{R}^{(2\ell+1)\times(2\ell+1)}$. If each angular order contains $C$ feature channels, the complete matrix acting on the direct sum of all compact irreducible coordinates is
\begin{align}
\mathbf{D}_{C}(\mathbf{Q})=\bigoplus_{\ell=0}^{L}\left[\mathbf{I}_{C}\otimes\mathbf{D}^{(\ell)}(\mathbf{Q})\right],\quad \mathbf{D}_{C}(\mathbf{Q})^{\top}\mathbf{D}_{C}(\mathbf{Q})=\mathbf{I}.
\label{eq:aiden-block-representation}
\end{align}
Here, $L$ is the maximum angular order, $\mathbf{I}_{C}$ is the $C\times C$ identity matrix, $\otimes$ denotes the Kronecker product, and $\bigoplus$ denotes a block-diagonal direct sum. If $\mathbf{z}_i$ stacks the independent irreducible coordinates of all angular components of atom $i$, equivariance is written compactly as $\mathbf{z}_i(g\mathcal{X})=\mathbf{D}_{C}(\mathbf{Q})\mathbf{z}_i(\mathcal{X})$. In the Cartesian tensor representation used by the encoder, the equivalent statement is $\mathbf{h}_i^{(\ell)}(g\mathcal{X})=\mathscr{D}_{\ell}(\mathbf{Q})\mathbf{h}_i^{(\ell)}(\mathcal{X})$.

Spatial inversion corresponds to $\mathbf{Q}=-\mathbf{I}$. A rank-$\ell$ angular basis acquires the parity factor $(-1)^{\ell}$, so its compact matrix representation is
\begin{align}
\boldsymbol{\Pi}_{C}=\bigoplus_{\ell=0}^{L}\left[(-1)^{\ell}\mathbf{I}_{C(2\ell+1)}\right],\quad \mathbf{D}^{(\ell)}(-\mathbf{I})=(-1)^{\ell}\mathbf{I}_{2\ell+1}.
\label{eq:aiden-parity-matrix}
\end{align}
Here, $\boldsymbol{\Pi}_{C}$ is the block-diagonal parity matrix acting on all angular channels. The analytic Cartesian harmonic $\mathbf{Y}^{(\ell)}(\widehat{\mathbf{d}})$ is constructed from the unit direction $\widehat{\mathbf{d}}$ and projected onto the symmetric traceless rank-$\ell$ subspace by $\mathcal{P}_{\ell}^{\mathrm{irr}}$. The operator $\mathcal{P}_{\ell}^{\mathrm{irr}}$ symmetrizes the tensor and removes its traces. Since an orthogonal transformation preserves the Euclidean metric, $\sum_aQ_{ab}Q_{ac}=\delta_{bc}$, where $\delta_{bc}$ is the Kronecker delta, this projection commutes with rotations. Therefore
\begin{align}
\mathbf{Y}^{(\ell)}(\widehat{\mathbf{d}}\mathbf{Q}^{\top})=\mathscr{D}_{\ell}(\mathbf{Q})\mathbf{Y}^{(\ell)}(\widehat{\mathbf{d}}),\quad \mathcal{P}_{\ell}^{\mathrm{irr}}\mathscr{D}_{\ell}(\mathbf{Q})=\mathscr{D}_{\ell}(\mathbf{Q})\mathcal{P}_{\ell}^{\mathrm{irr}}.
\label{eq:aiden-irrep-transform}
\end{align}

The coupling $\mathcal{C}_{\ell_1,\ell_2}^{\ell}$ combines rank-$\ell_1$ and rank-$\ell_2$ irreducible Cartesian tensors into their allowed rank-$\ell$ component. It is built from tensor products, contractions with the Euclidean metric, and irreducible projection, and therefore satisfies the intertwining relation
\begin{align}
\mathcal{C}_{\ell_1,\ell_2}^{\ell}\left(\mathscr{D}_{\ell_1}(\mathbf{Q})\mathbf{T}_1,\mathscr{D}_{\ell_2}(\mathbf{Q})\mathbf{T}_2\right)=\mathscr{D}_{\ell}(\mathbf{Q})\mathcal{C}_{\ell_1,\ell_2}^{\ell}(\mathbf{T}_1,\mathbf{T}_2).
\label{eq:aiden-coupling-equivariance}
\end{align}
Here, $\mathbf{T}_1$ and $\mathbf{T}_2$ are input irreducible tensors. The order-$\nu$ Cartesian ACE polynomial $\mathcal{B}_{\nu}^{(\ell)}$ is recursively assembled from these couplings, where $\nu$ denotes the correlation order. Since all learned channel maps act only between channels of the same angular order, induction over $\nu$ gives $\mathcal{B}_{\nu}^{(\ell)}[\mathscr{D}(\mathbf{Q})\mathbf{T}]=\mathscr{D}_{\ell}(\mathbf{Q})\mathcal{B}_{\nu}^{(\ell)}[\mathbf{T}]$, with $\mathscr{D}(\mathbf{Q})=\bigoplus_{\ell=0}^{L}\mathscr{D}_{\ell}(\mathbf{Q})$ denoting the direct action on all angular orders.

For a directed periodic edge $e=(j,\mathbf{n}\to i)$, Eq.~\ref{eq:atomic-edge} defines the displacement $\mathbf{d}_e$, distance $r_e=\|\mathbf{d}_e\|_2$, and unit direction $\widehat{\mathbf{d}}_e=\mathbf{d}_e/r_e$. Under the global transformation, $\mathbf{d}_e'=\mathbf{d}_e\mathbf{Q}^{\top}$, $r_e'=r_e$, and $\widehat{\mathbf{d}}_e'=\widehat{\mathbf{d}}_e\mathbf{Q}^{\top}$. The radial representation $\mathbf{b}(r_e)$ depends only on $r_e$ and is therefore invariant. The distance-defined edge set $\mathcal{N}_i$ is unchanged, and the elemental input $\mathbf{x}_i$ does not depend on spatial orientation. In GIE, every learned amplitude multiplying $\mathbf{Y}^{(\ell)}(\widehat{\mathbf{d}}_e)$ is determined from these scalar quantities. The scalar branch is invariant and the higher-order branches follow Eq.~\ref{eq:aiden-irrep-transform}, giving
\begin{align}
\mathbf{h}_i^{(0,\ell)}(g\mathcal{X})=\mathscr{D}_{\ell}(\mathbf{Q})\mathbf{h}_i^{(0,\ell)}(\mathcal{X}),\quad 0\le\ell\le L.
\label{eq:aiden-gie-equivariance}
\end{align}
Here, $\mathbf{h}_i^{(0,\ell)}$ is the rank-$\ell$ feature produced by GIE before the ACE interaction. Every channel map $\mathcal{L}^{(\ell)}$ acts only within fixed $\ell$ and identically on its Cartesian components, so $\mathcal{L}^{(\ell)}\mathscr{D}_{\ell}(\mathbf{Q})=\mathscr{D}_{\ell}(\mathbf{Q})\mathcal{L}^{(\ell)}$. Bias terms are restricted to $\ell=0$. The normalization $\mathcal{V}_{\ell}$ is also equivariant because its denominator is constructed from rotationally invariant sums of squared Cartesian components and its learned gain acts only on channels. Thus $\mathcal{V}_{\ell}[\mathscr{D}_{\ell}(\mathbf{Q})\mathbf{h}]=\mathscr{D}_{\ell}(\mathbf{Q})\mathcal{V}_{\ell}[\mathbf{h}]$ for any rank-$\ell$ feature tensor $\mathbf{h}$.

For Cartesian ACE, $\mathbf{p}_e^{(\ell)}$ denotes the rank-$\ell$ edge tensor in Eq.~\ref{eq:ace-edge-field}, while $\mathbf{w}_{e,\ell_1\ell_2\ell}$ is its learned radial channel weight. Since $\mathbf{w}_{e,\ell_1\ell_2\ell}$ depends only on $\mathbf{b}(r_e)$, it is invariant. The source feature and Cartesian harmonic transform under $\mathscr{D}_{\ell_1}(\mathbf{Q})$ and $\mathscr{D}_{\ell_2}(\mathbf{Q})$, respectively. Equation~\ref{eq:aiden-coupling-equivariance} therefore gives $\mathbf{p}_e^{(\ell)}(g\mathcal{X})=\mathscr{D}_{\ell}(\mathbf{Q})\mathbf{p}_e^{(\ell)}(\mathcal{X})$. In particular, the scalar component $\mathbf{p}_e^{(0)}$ is invariant. The neighbor coefficient $a_e$ is constructed from invariant edge information and is also unchanged. The aggregated ACE field $\boldsymbol{\Xi}_i^{(\ell)}$ and the corresponding ACE output therefore satisfy
\begin{align}
\boldsymbol{\Xi}_i^{(\ell)}(g\mathcal{X})=\mathscr{D}_{\ell}(\mathbf{Q})\boldsymbol{\Xi}_i^{(\ell)}(\mathcal{X}),\quad \mathbf{h}_i^{(1,\ell)}(g\mathcal{X})=\mathscr{D}_{\ell}(\mathbf{Q})\mathbf{h}_i^{(1,\ell)}(\mathcal{X}),\quad 0\le\ell\le L.
\label{eq:aiden-ace-equivariance}
\end{align}
Here, $\boldsymbol{\Xi}_i^{(\ell)}$ is the rank-$\ell$ component of the aggregated ACE field and $\mathbf{h}_i^{(1,\ell)}$ is the ACE output. The gate $\sigma_0\mathcal{F}_{\mathrm{g}}[\boldsymbol{\Xi}_i^{(0)}]$ is generated only from the invariant $\ell=0$ sector, where $\sigma_0$ is the sigmoid and $\mathcal{F}_{\mathrm{g}}$ is the learned gating map. It therefore acts as a rotationally invariant channel-wise scalar on every angular component.

\textbf{SO(3) equivariance of TECE and radial rotary attention.} TECE converts the global irreducible Cartesian representation into real-spherical coordinates and then rotates these coordinates into an edge-aligned local frame. Let $\mathbf{S}^{(\ell)}(\mathbf{Q})$ denote the real-spherical matrix representation of the rank-$\ell$ irrep. With $C$ channel copies, we define $\mathbf{S}_{C}(\mathbf{Q})=\bigoplus_{\ell=0}^{L}[\mathbf{I}_{C}\otimes\mathbf{S}^{(\ell)}(\mathbf{Q})]$. The fixed orthogonal map $\mathcal{U}$ converts compact Cartesian irrep coordinates into the corresponding real-spherical coordinates. Since the Cartesian and real-spherical bases represent the same rotation, $\mathcal{U}$ satisfies
\begin{align}
\mathcal{U}\mathbf{D}_{C}(\mathbf{Q})=\mathbf{S}_{C}(\mathbf{Q})\mathcal{U},\quad \mathcal{U}^{\dagger}\mathbf{S}_{C}(\mathbf{Q})=\mathbf{D}_{C}(\mathbf{Q})\mathcal{U}^{\dagger}.
\label{eq:aiden-intertwiner}
\end{align}
Here, $\mathcal{U}^{\dagger}$ is the inverse basis transformation; because $\mathcal{U}$ is orthogonal in the real basis, $\mathcal{U}^{\dagger}=\mathcal{U}^{\top}$. For each edge $e$, let $\mathbf{G}_e\in\mathrm{SO}(3)$ denote the spatial rotation that aligns the local $y$ axis with $\widehat{\mathbf{d}}_e$. Its action on the spherical feature representation is $\mathbf{D}_e=\mathbf{S}_{C}(\mathbf{G}_e)$. After the global rotation $\mathbf{Q}$, the transformed edge direction defines another frame $\mathbf{G}_e'$ and corresponding feature-space matrix $\mathbf{D}_e'=\mathbf{S}_{C}(\mathbf{G}_e')$. Since both frames align the same physical edge with the local $y$ axis, they can differ only by a residual rotation around that axis. We define this rotation as $\mathbf{H}_e=\mathbf{G}_e'\mathbf{Q}\mathbf{G}_e^{\top}\in\mathrm{SO}(2)_y$ and its feature-space representation as $\mathbf{S}_e=\mathbf{S}_{C}(\mathbf{H}_e)$. The global and local transformations are therefore related by
\begin{align}
\mathbf{D}_e'\mathbf{S}_{C}(\mathbf{Q})=\mathbf{S}_e\mathbf{D}_e.
\label{eq:aiden-frame-relation}
\end{align}
Here, $\mathrm{SO}(2)_y$ denotes the subgroup of rotations around the local $y$ axis. If $\mathbf{z}_e^{\mathrm{s}}$ and $\mathbf{z}_e^{\mathrm{t}}$ denote the edge-local spherical source and target features, respectively, Eq.~\ref{eq:aiden-frame-relation} immediately gives $\mathbf{z}_e^{\mathrm{s}\prime}=\mathbf{S}_e\mathbf{z}_e^{\mathrm{s}}$ and $\mathbf{z}_e^{\mathrm{t}\prime}=\mathbf{S}_e\mathbf{z}_e^{\mathrm{t}}$.

For magnetic order $m>0$, the two real components at fixed $(\ell,m)$ can be represented by the complex quantity $z_{\ell m}=z_{\ell m}^{\mathrm{Re}}+\mathrm{i}z_{\ell m}^{\mathrm{Im}}$, where $\mathrm{i}^{2}=-1$. If $\alpha_e$ is the angle of the residual rotation $\mathbf{H}_e$, this component transforms as $z_{\ell m}\mapsto\exp(\mathrm{i}m\alpha_e)z_{\ell m}$, while the $m=0$ sector is invariant. The radial operator $\boldsymbol{\Omega}_e[\mathbf{b}(r_e)]$ depends only on the invariant radial vector and applies the same coefficient to the two real components belonging to a fixed $m$, so it commutes with $\mathbf{S}_e$. The learned $\mathrm{SO}(2)$ channel maps have the same property.

The second-order local correlator $\mathcal{E}$ combines magnetic frequencies according to the usual addition rule. For components with orders $m_1$ and $m_2$, the product $z_{m_1}z_{m_2}$ transforms with frequency $m_1+m_2$, while $z_{m_1}\overline{z}_{m_2}$ transforms with frequency $m_1-m_2$, where the overline denotes complex conjugation. The learned coefficients of $\mathcal{E}$ are generated from the invariant $m=0$ sector and radial quantities and therefore behave as scalars under the residual rotation. Consequently, $\mathcal{E}(\mathbf{S}_e\mathbf{z})=\mathbf{S}_e\mathcal{E}(\mathbf{z})$, where $\mathbf{z}$ denotes the local source-target feature collection entering the correlator.

Radial rotary attention assigns an invariant scalar weight to each edge and attention head. For angular order $\ell$, magnetic order $m$, and head $h$, let $\mathbf{q}_{e,\ell mh}\in\mathbb{C}^{C_h}$ and $\mathbf{k}_{e,\ell mh}\in\mathbb{C}^{C_h}$ denote the projected query and key vectors, where $C_h$ is the number of channels in head $h$. Under the residual rotation, $\mathbf{q}_{e,\ell mh}'=\exp(\mathrm{i}m\alpha_e)\mathbf{q}_{e,\ell mh}$ and $\mathbf{k}_{e,\ell mh}'=\exp(\mathrm{i}m\alpha_e)\mathbf{k}_{e,\ell mh}$. The radial phase $\varphi_{eh}$ depends only on invariant radial features. The Hermitian channel contraction is defined by $\langle\mathbf{u},\mathbf{v}\rangle_{\mathrm{ch}}=\sum_{c=1}^{C_h}u_c^{*}v_c$, where $c$ is the channel index and $u_c^{*}$ denotes complex conjugation. The common residual phase therefore cancels exactly,
\begin{align}
\mathrm{Re}\left\langle\exp(\mathrm{i}m\alpha_e)\mathbf{q}_{e,\ell mh},\exp(\mathrm{i}m\varphi_{eh})\exp(\mathrm{i}m\alpha_e)\mathbf{k}_{e,\ell mh}\right\rangle_{\mathrm{ch}}=\mathrm{Re}\left\langle\mathbf{q}_{e,\ell mh},\exp(\mathrm{i}m\varphi_{eh})\mathbf{k}_{e,\ell mh}\right\rangle_{\mathrm{ch}}.
\label{eq:aiden-rra-invariance}
\end{align}
Here, $\mathrm{Re}$ extracts the real part. It follows that the score $\xi_{eh}$ in Eq.~\ref{eq:rra-score} and the cutoff-weighted softmax coefficient $a_{eh}$ are invariant. The matrix $\mathbf{A}_e$ applies these scalar coefficients to the channel blocks of the attention heads and therefore acts only on channel indices. It commutes with $\mathbf{S}_e$.

Let $\mathbf{v}_e$ denote the edge-local output after radial modulation, the correlator $\mathcal{E}$, and RRA. The previous results give $\mathbf{v}_e'=\mathbf{S}_e\mathbf{v}_e$. From Eq.~\ref{eq:aiden-frame-relation}, $\mathbf{D}_e'^{\dagger}\mathbf{S}_e=\mathbf{S}_{C}(\mathbf{Q})\mathbf{D}_e^{\dagger}$. Applying the inverse frame rotation and the inverse spherical-to-Cartesian basis transformation therefore gives
\begin{align}
\mathcal{U}^{\dagger}\mathbf{D}_e'^{\dagger}\mathbf{v}_e'=\mathbf{D}_{C}(\mathbf{Q})\mathcal{U}^{\dagger}\mathbf{D}_e^{\dagger}\mathbf{v}_e.
\label{eq:aiden-tece-message}
\end{align}
Here, $\mathbf{D}_e^{\dagger}$ returns an edge-local spherical feature to the global spherical frame, and $\mathcal{U}^{\dagger}$ maps the spherical representation back to the irreducible Cartesian representation. Neighbor summation is linear, the residual connection adds quantities carrying the same representation, and the final normalization is equivariant. Hence the final encoder output satisfies
\begin{align}
\mathbf{h}_i^{\star(\ell)}(g\mathcal{X})=\mathscr{D}_{\ell}(\mathbf{Q})\mathbf{h}_i^{\star(\ell)}(\mathcal{X}),\quad 0\le\ell\le L,\quad \mathbf{Q}\in\mathrm{SO}(3).
\label{eq:aiden-encoder-equivariance}
\end{align}
Thus, the implemented encoder is analytically equivariant under proper rotations and invariant to global translations.

\textbf{Continuous density covariance and lattice periodicity.} The decoder maps the final rank-$\ell$ atomic representation to the coefficient tensor $\mathbf{c}_{ikp}^{\chi(\ell)}=\mathcal{L}_{kp}^{\chi(\ell)}\mathbf{h}_i^{\star(\ell)}+\delta_{\ell0}\mu_{kp}^{\chi}$. Here, $i$ is the atom index, $k$ labels the environmental field channel, $p$ labels the Gaussian radial basis function, and $\chi\in\{\mathrm{L},\mathrm{R}\}$ denotes the left or right decoder branch. The map $\mathcal{L}_{kp}^{\chi(\ell)}$ acts only on channels within fixed $\ell$, $\delta_{\ell0}$ is the Kronecker delta, and $\mu_{kp}^{\chi}$ is a scalar bias present only for $\ell=0$. The coefficient tensor therefore transforms as $\mathbf{c}_{ikp}^{\chi(\ell)}(g\mathcal{X})=\mathscr{D}_{\ell}(\mathbf{Q})\mathbf{c}_{ikp}^{\chi(\ell)}(\mathcal{X})$.

For a periodic image $\eta=(i,\mathbf{n})$ contributing to a query point $\mathbf{r}$, Eq.~\ref{eq:query-images} defines $\boldsymbol{\delta}_{\eta}=\mathbf{r}-(\mathbf{R}_i+\mathbf{n}\mathbf{A})$, $r_{\eta}=\|\boldsymbol{\delta}_{\eta}\|_2$, and $\widehat{\boldsymbol{\delta}}_{\eta}=\boldsymbol{\delta}_{\eta}/r_{\eta}$ for $r_{\eta}>0$. Under the simultaneous transformation of the structure and query point, $\boldsymbol{\delta}_{\eta}'=\boldsymbol{\delta}_{\eta}\mathbf{Q}^{\top}$, $r_{\eta}'=r_{\eta}$, and $\widehat{\boldsymbol{\delta}}_{\eta}'=\widehat{\boldsymbol{\delta}}_{\eta}\mathbf{Q}^{\top}$. The radial basis $\widetilde{R}_{\ell p}(r_{\eta})$ is therefore invariant, while the angular basis transforms according to Eq.~\ref{eq:aiden-irrep-transform}. The final scalar contraction can be seen most clearly in the compact irreducible basis. Let $\mathbf{c}$ and $\mathbf{y}$ denote the coordinate vectors of the rank-$\ell$ decoder coefficient and angular harmonic. Both transform with the same orthogonal representation matrix $\mathbf{D}^{(\ell)}(\mathbf{Q})$, so we have 
\begin{align}
\left[\mathbf{D}^{(\ell)}(\mathbf{Q})\mathbf{c}\right]^{\top}\left[\mathbf{D}^{(\ell)}(\mathbf{Q})\mathbf{y}\right]=\mathbf{c}^{\top}\mathbf{D}^{(\ell)}(\mathbf{Q})^{\top}\mathbf{D}^{(\ell)}(\mathbf{Q})\mathbf{y}=\mathbf{c}^{\top}\mathbf{y}.
\label{eq:aiden-decoder-matrix-contraction}
\end{align}
The transpose $^{\top}$ acts on the compact coordinate vector, and the last equality follows from the orthogonality of $\mathbf{D}^{(\ell)}(\mathbf{Q})$. This matrix contraction is equivalent to the Frobenius contraction between the corresponding Cartesian coefficient tensor and Cartesian harmonic. Every term entering the environment-dependent field $\Phi_k^{\chi}$ is therefore invariant under the joint transformation of the structure and query point.

The one-center field $\phi_{\eta k_0}^{\chi}$ depends only on the element $Z_i$ and the scalar radial functions $\widetilde{R}_{0p}(r_{\eta})$ and is also invariant. Here, $k_0$ labels the one-center field channel. Consequently,
\begin{align}
\rho_{\mathrm{init}}(g\mathbf{r};g\mathcal{X})=\rho_{\mathrm{init}}(\mathbf{r};\mathcal{X}),\quad \rho_{\mathrm{env}}(g\mathbf{r};g\mathcal{X})=\rho_{\mathrm{env}}(\mathbf{r};\mathcal{X}),\quad \widehat{\rho}_{\boldsymbol{\theta}}(g\mathbf{r};g\mathcal{X})=\widehat{\rho}_{\boldsymbol{\theta}}(\mathbf{r};\mathcal{X}),
\label{eq:aiden-density-covariance}
\end{align}
for $\mathbf{Q}\in\mathrm{SO}(3)$. Here, $\rho_{\mathrm{init}}$ is the element-dependent one-center density, $\rho_{\mathrm{env}}$ is the environment-dependent contribution, and $\widehat{\rho}_{\boldsymbol{\theta}}=\rho_{\mathrm{init}}+\rho_{\mathrm{env}}$ is the complete predicted density with model parameters $\boldsymbol{\theta}$. At the angular-basis level, inversion gives $\mathbf{Y}^{(\ell)}(-\widehat{\mathbf{d}})=(-1)^{\ell}\mathbf{Y}^{(\ell)}(\widehat{\mathbf{d}})$ and is represented by $\boldsymbol{\Pi}_{C}$. The scalar decoder contraction is therefore compatible with the corresponding parity transformation of its coefficient tensor. The complete implemented mapping under inversion is tested directly below. Lattice periodicity follows independently from rotational equivariance. Let $\mathbf{m}\in\mathbb{Z}^{3}$ denote an arbitrary lattice translation index. The image $\eta=(i,\mathbf{n})$ contributing at $\mathbf{r}$ can be paired with $\eta'=(i,\mathbf{n}+\mathbf{m})$ at $\mathbf{r}+\mathbf{m}\mathbf{A}$. Their relative displacements satisfy $\boldsymbol{\delta}_{\eta'}(\mathbf{r}+\mathbf{m}\mathbf{A})=\boldsymbol{\delta}_{\eta}(\mathbf{r})$. Thus translating the query point by a lattice vector only relabels the periodic images in $\mathcal{M}(\mathbf{r})$, and
\begin{align}
\widehat{\rho}_{\boldsymbol{\theta}}(\mathbf{r}+\mathbf{m}\mathbf{A};\mathcal{X})=\widehat{\rho}_{\boldsymbol{\theta}}(\mathbf{r};\mathcal{X}).
\label{eq:aiden-periodicity}
\end{align}

\begin{figure}[t]
    \centering
    \includegraphics[width=\linewidth]{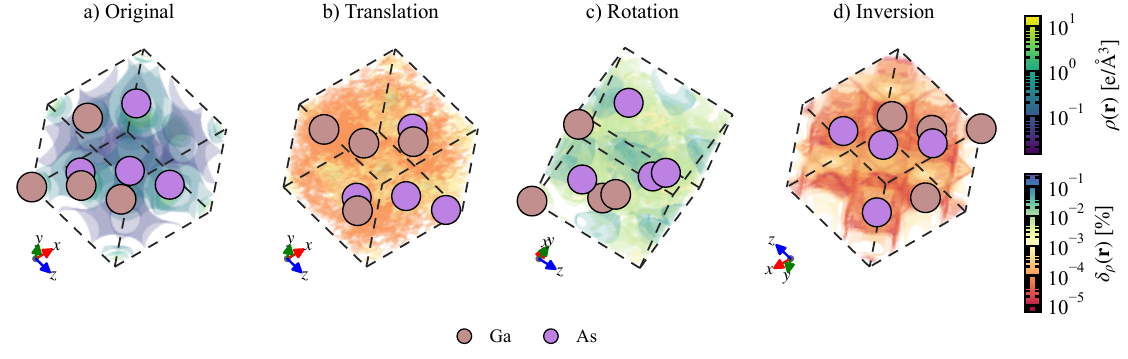}
    \caption{\textbf{E(3)-equivariance test of the model-predicted charge density for GaAs.} (a) Charge density $\rho(\mathbf{r})$ predicted by AIDEN for zincblende GaAs (\href{https://next-gen.materialsproject.org/materials/mp-2534/}{\texttt{mp-2534}}) on a $32\times32\times32$ grid. (b)-(d) Normalized pointwise deviation $\delta_{\rho}(\mathbf{r}):=|\rho_g(\mathbf{r})-\rho(\mathbf{r})|/\rho(\mathbf{r})$, where $\rho_g$ denotes the prediction after applying the transformation $g$ and $\rho(\mathbf{r})$ denotes the reference charge density at $\mathbf{r}\in \mathbb{R}^3$ in (a). The tested transformations are (b) translation by $\mathbf{t}=(3,-5,7)/32$, (c) proper rotation about $\mathbf{n}=(-1,-2,1)/\sqrt{6}$ with $\det\mathbf{R}=+1$, and (d) inversion $\mathbf{r}\rightarrow-\mathbf{r}$ with $\det\mathbf{R}=-1$. The transformed predictions remain numerically consistent with the reference. Notably, inversion is not a symmetry operation of non-centrosymmetric zincblende GaAs, providing a direct test beyond the crystal space-group symmetries.}
    \label{fig:eq-gaas}
\end{figure}

\textbf{Numerical verification of the complete E(3) transformation.} We further evaluate the complete trained mapping by transforming the \texttt{.cif} geometry and querying the density at the correspondingly transformed spatial coordinates. Fig.~\ref{fig:eq-gaas} uses non-centrosymmetric zinc-blende GaAs and compares the original prediction with predictions after a global translation, a generic proper rotation, and spatial inversion. The normalized $L_1$ deviations are $7.72\times10^{-6}$, $7.27\times10^{-5}$, and $9.51\times10^{-7}$, respectively. The inversion test is particularly useful because inversion is not a crystal symmetry of zinc-blende GaAs. The inverted structure is related to the original one by a Euclidean transformation but is not mapped back to the original \texttt{.cif} by a space-group operation. The measured deviation therefore probes the structure-to-field transformation law itself rather than an intrinsic symmetry of the selected crystal. Together with the exact translation invariance and the analytical $\mathrm{SO}(3)$ derivation above, the inversion experiment numerically verifies the $\mathrm{O}(3)$ extension of the implemented mapping within the measured tolerance. Since any improper orthogonal transformation can be decomposed into inversion and a proper rotation, AIDEN realizes an $\mathrm{E}(3)$-consistent and lattice-periodic mapping from periodic atomic structures to continuous scalar charge-density fields.

\section{Detailed implementations of AIDEN}
\label{app:aiden-details}

This section collects the implementation details needed to reproduce the AIDEN architecture in Sec.~\ref{sec:model}. We retain only the constructions that define the periodic geometric basis, the two equivariant cluster-expansion interactions, and the continuous GTO decoder; numerical bookkeeping that does not affect the model definition is omitted.

\subsection{Periodic graph and geometric basis}
\label{app:graph-basis}

The elemental descriptor $\mathbf q(Z)$ contains atomic number, period, group, total and orbital-resolved valence counts, electronegativity, atomic and covalent radii, first ionization energy, electron affinity, dipole polarizability, and GS magnetic moment. It is concatenated with the one-hot vector $\mathbf o(Z)$ and feature-wise min-max normalized over the 118 elements, yielding the $133$-dimensional input $\mathbf x_i$ used in Sec.~\ref{sec:encoder}. No structure-dependent information enters this elemental representation.

For periodic crystals, all atomic images satisfying the cutoff in Eq.~\ref{eq:atomic-edge} are enumerated before the nearest $N_{\rm nbr}$ images are retained for each target atom. We use $N_{\rm nbr}=100$ and $r_{\rm at}=4\,\text{\AA}$. Query-atom images are constructed independently with the orbital cutoff $r_{\rm orb}$ in Eq.~\ref{eq:query-images} and are not neighbor-capped. For VASP CHGCAR data, FFT-grid points are interpreted in fractional coordinates and mapped to Cartesian positions by $\mathbf r_g=\mathbf u_g\mathbf A$; the stored density values are divided by $|\det\mathbf A|$ to obtain the supervised density in $e/\text{\AA}^3$.

The atomic edge basis consists of a zero-order spherical-Bessel radial expansion multiplied by the standard compact polynomial envelope $f_{\rm c}$~\citep{DimeNet}, together with irreducible Cartesian harmonics,
\begin{align}
b_b(r)&=\sqrt{\frac{2}{r_{\rm at}}}\frac{\sin(b\pi r/r_{\rm at})}{r}f_{\rm c}(r),\quad b=1,\dots,N_{\rm B},\label{eq:bessel-basis}\\
\mathbf Y^{(\ell)}(\widehat{\mathbf v})&=\frac{(2\ell-1)!!}{\ell!}\mathcal P_{\ell}^{\rm irr}\left(\widehat{\mathbf v}^{\otimes\ell}\right),\quad \mathbf Y^{(0)}=1.\label{eq:cart-harmonic}
\end{align}
Here $\mathcal P_{\ell}^{\rm irr}$ denotes the symmetric-traceless projection used throughout the Cartesian encoder. We use $N_{\rm B}=8$. The same radial vector $\mathbf b(r_e)$ is shared by GIE, Cartesian ACE, and TECE.

\subsection{Cartesian atomic cluster expansion}
\label{app:ace-detail}

For an angular path $(\ell_1,\ell_2,\ell)$, define $\kappa=(\ell_1+\ell_2-\ell)/2$. The Cartesian coupling contracts $\kappa$ index pairs and projects the remaining tensor to rank $\ell$,
\begin{align}
\mathcal C_{\ell_1,\ell_2}^{\ell}(\mathbf T_1,\mathbf T_2)&=3^{-\kappa/2}\mathcal P_{\ell}^{\rm irr}\left(\mathcal K_{\kappa}(\mathbf T_1,\mathbf T_2)\right),\label{eq:cart-coupling}
\end{align}
\begin{align}
\mathcal T_L&=\left\{(\ell_1,\ell_2,\ell)\in\{0,\dots,L\}^3\ \middle|\ |\ell_1-\ell_2|\le\ell\le\min(L,\ell_1+\ell_2),\ \ell_1+\ell_2-\ell\equiv0\pmod2\right\}.\label{eq:ace-paths}
\end{align}
This is the coupling used in Eq.~\ref{eq:ace-edge-field}; for $L=3$, the path set contains $23$ admissible triples.

The equivariant RMS normalization applied before the interactions and at the encoder output rescales each angular order by a rotation-invariant channel RMS,
\begin{align}
\mathcal V_{\ell}[\mathbf h_i^{(\ell)}]_c=\frac{\gamma_{\ell c}\mathbf h_{ic}^{(\ell)}}{\sqrt{\max\left(C^{-1}\sum_{c'=1}^{C}\|\mathbf h_{ic'}^{(\ell)}\|_{\rm F}^2,\epsilon\right)}}.
\label{eq:equivariant-rmsnorm}
\end{align}
The learned gain $\gamma_{\ell c}$ acts only on channels, so the Cartesian transformation law is unchanged. For the atomic aggregation in Eq.~\ref{eq:ace-compact-update}, the scalar sector of the edge field defines an invariant logit $s_e=\mathcal F_{\rm a}[\mathbf p_e^{(0)}\oplus\mathbf b(r_e)]$. The corresponding destination-wise coefficient is
\begin{align}
a_e=\frac{f_{\rm c}(r_e){\rm e}^{s_e}}{\sum_{e'\in\mathcal N_i}f_{\rm c}(r_{e'}){\rm e}^{s_{e'}}}.
\label{eq:ace-attention-appendix}
\end{align}
The same scalar $a_e$ multiplies every angular order of edge $e$. 

To make the correlation polynomial in Eq.~\ref{eq:ace-compact-update} explicit, let $\boldsymbol\Upsilon_i=\{\mathbf1_C+\sigma_0\mathcal F_{\rm g}[\boldsymbol\Xi_i^{(0)}]\}\odot\boldsymbol\Xi_i$. Starting from $\mathbf B_i^{[1](\ell)}=\boldsymbol\Upsilon_i^{(\ell)}$, higher correlation orders are generated recursively and combined as
\begin{align}
\mathbf B_i^{[q](\ell)}&=\sum_{(\ell_1,\ell_2,\ell)\in\mathcal T_L}\mathcal C_{\ell_1,\ell_2}^{\ell}\left(\mathbf B_i^{[q-1](\ell_1)},\boldsymbol\Upsilon_i^{(\ell_2)}\right),\quad \mathcal B_{\nu}^{(\ell)}(\boldsymbol\Upsilon_i)=\sum_{q=1}^{\nu}\mathcal L_{\rm corr,q}^{(\ell)}\mathbf B_i^{[q](\ell)}.
\label{eq:ace-polynomial}
\end{align}
We use $\nu=2$. This order refers to the explicit Cartesian correlation inside the ACE block; the preceding GIE representation already contains one-hop environmental information.

\subsection{TECE and RRA}
\label{app:tece-detail}
\label{app:attention}

The orthogonal map $\mathcal U$ converts each irreducible Cartesian tensor into its real-spherical components, after which $\mathbf D_e$ aligns the local $y$ axis with $\widehat{\mathbf d}_e$. For $|m|\le M$, the compact representation contains $D_M=(L+1)+2\sum_{m=1}^{M}(L+1-m)$ real components. With $L=M=3$, all magnetic components are retained and $D_M=16$. The source and destination representations are modulated by radial weights before entering the edge correlator. Suppressing channel indices, this operation can be written consistently with Eq.~\ref{eq:tece-compact-update} as
\begin{align}
\mathbf z_e=\boldsymbol\Omega_e[\mathbf b(r_e)]\odot\left(\mathbf D_e\mathcal U\overline{\mathbf h}_j^{(1)}\oplus\mathbf D_e\mathcal U\overline{\mathbf h}_i^{(1)}\right),
\label{eq:tece-radial}
\end{align}
where the same radial coefficient is applied to the real-imaginary pair associated with a fixed nonzero $m$. This preserves the local SO(2) transformation law. The operator $\mathcal E$ combines a direct branch, an SO(2)-gated branch, and a second-order product branch. Absorbing the fixed-frequency channel projections into the corresponding operators, its structure is
\begin{align}
\mathcal E(\mathbf z_e)=\mathcal L_{\rm out}^{\rm TECE}\left[\frac{\mathbf v_e+\mathcal G(\mathbf v_e)+\Pi^{(2)}(\mathbf v_e,\mathbf w_e)}{\sqrt3}\right],\quad \mathbf v_e=\mathcal L_{
m v}\mathbf z_e,\quad \mathbf w_e=\mathcal L_{
m w}\mathbf z_e.
\label{eq:tece-edge-correlator}
\end{align}
For a complex local mode $z_m$, a residual rotation around the edge axis gives $z_m\mapsto{\rm e}^{{\rm i}m\vartheta}z_m$. Accordingly, $\Pi^{(2)}$ retains only products with output frequencies $m=m_1+m_2$ or $m=|m_1-m_2|$. Their coefficients are generated from the invariant $m=0$ sector, so the product branch remains SO(2)-equivariant. The direct, gated, and product branches are then returned to the global Cartesian representation through $\mathbf D_e^\dagger$ and $\mathcal U^\dagger$ as in Eq.~\ref{eq:tece-compact-update}.

RRA is evaluated from the unmodulated local target and source tensors. Splitting the $C_{\rm e}$ channels into $H$ heads gives $C_h=C_{\rm e}/H$; for $m>0$, each real-imaginary pair is identified with a complex vector, and the head-wise contraction is $\langle\mathbf u,\mathbf v\rangle_{\rm ch}=\sum_{c=1}^{C_h}u_c^*v_c$. The score $\xi_{eh}$ is given in Eq.~\ref{eq:rra-score}. Its normalized coefficient is
\begin{align}
a_{eh}=\frac{f_{\rm c}(r_e){\rm e}^{\xi_{eh}}}{\sum_{e'\in\mathcal N_i}f_{\rm c}(r_{e'}){\rm e}^{\xi_{e'h}}},
\label{eq:rra-attention-appendix}
\end{align}
which is invariant because it depends only on equal-frequency inner products, radial quantities, and the cutoff envelope. We use $H=4$ and $C_{\rm e}=C=48$.

\subsection{Continuous GTO decoder}
\label{app:decoder-detail}

The decoder uses $N_{\rm G}$ even-tempered Gaussian radial functions for every angular order. Their exponents and corrected radial factors are
\begin{align}
\alpha_p=\alpha_{\min}\left(\frac{\alpha_{\max}}{\alpha_{\min}}\right)^{\frac{p-1}{N_{\rm G}-1}},\quad \widetilde R_{\ell p}(r)=\mathcal Z_{\ell p}r^\ell{\rm e}^{-\alpha_pr^2}\left[1+\zeta_{\ell p}(r)\right],\quad \mathcal Z_{\ell p}=\sqrt{\frac{2(2\alpha_p)^{\ell+3/2}}{\Gamma(\ell+3/2)}}.
\label{eq:gaussian-radial}
\end{align}
We use $N_{\rm G}=8$, $\alpha_{\min}=0.15\,\text{\AA}^{-2}$, and $\alpha_{\max}=256\,\text{\AA}^{-2}$. The element-independent correction $\zeta_{\ell p}(r)$ is predicted from $50$ Gaussian distance features by a small MLP whose final affine layer is initialized to zero, so the decoder starts from the analytic GTO family. The same corrected radial basis is used by the one-center and environment-dependent branches.

The coefficient tensors $\mathbf c_{ikp}^{\chi(\ell)}$ and the continuous fields $\phi_{\eta k_0}^{\chi}$ and $\Phi_k^{\chi}$ are defined directly in Sec.~\ref{sec:decoder}. Because $\mathcal U$ is orthogonal on the irreducible subspace, the Frobenius contraction in Eq.~\ref{eq:continuous-fields} is equivalent to contracting the corresponding real-spherical coefficients and harmonics, without introducing an explicit magnetic index in the decoder. We use environmental rank $K=8$, one-center rank $K_0=1$, and orbital cutoff $r_{\rm orb}=3\,\text{\AA}$. As in BOA, the implementation applies a smooth absolute value to the left factor of each low-rank product before multiplication; this scalar stabilization is omitted from Eq.~\ref{eq:density-master} because it does not alter the equivariant structure of the decoder. The localized GTO basis provides a continuous representation independent of the evaluation grid. The bilinear form in Eq.~\ref{eq:density-master} generates cross-center terms when expanded; with the scalar stabilization, it still couples contributions from different centers. The learned factors are auxiliary fields, not occupied KS orbitals. Likewise, the one-center/environment split is a modeling choice: joint optimization does not impose a unique atomic partition or require the environment-dependent term to integrate to zero.

\subsection{Forward propagation}
\label{app:forward}

Given a structure $\mathcal X$, the equivariant encoder is evaluated once to obtain the atomic coefficients $\mathbf c_{ikp}^{\chi(\ell)}$, after which the continuous decoder evaluates $\widehat\rho_{\boldsymbol\theta}$ at arbitrary query positions. Alg.~\ref{alg:aiden-forward} summarizes the resulting forward propagation.

\begin{algorithm}[h]
\caption{Forward propagation of AIDEN}
\label{alg:aiden-forward}
\begin{algorithmic}[1]
\Require Structure $\mathcal X$ and query positions $\{\mathbf r_g\}_{g=1}^{N_{\rm q}}$
\Ensure $\{\widehat\rho_{\boldsymbol\theta}(\mathbf r_g;\mathcal X)\}_{g=1}^{N_{\rm q}}$

\State Construct $\{\mathbf x_i\}_i$ and the periodic graph $\{\mathcal N_i\}_i$ with $\{\mathbf b(r_e),\mathbf Y^{(\ell)}(\widehat{\mathbf d}_e)\}_e$ from Eq.~\ref{eq:atomic-edge}.
\State $\{\mathbf h_i^{(0,\ell)}\}_{\ell=0}^{L}\leftarrow{\rm GIE}(\mathcal X)$ using Eqs.~\ref{eq:gie-scalar}-\ref{eq:gie-compact}.
\State $\mathbf h_i^{(1,\ell)}\leftarrow{\rm ACE}\!\left(\{\mathcal V_{\ell'}[\mathbf h_i^{(0,\ell')}]\}_{\ell'=0}^{L}\right)$ using Eq.~\ref{eq:ace-compact-update}.
\State $\mathbf h_i^{(2)}\leftarrow{\rm TECE}\!\left(\{\mathcal V_{\ell}[\mathbf h_i^{(1,\ell)}]\}_{\ell=0}^{L}\right)$ using Eq.~\ref{eq:tece-compact-update}.
\State $\mathbf h_i^{\star(\ell)}\leftarrow\mathcal V_\ell[\mathbf h_i^{(2,\ell)}]$ using Eq.~\ref{eq:tece-output-normalization}.
\State $\mathbf c_{ikp}^{\chi(\ell)}\leftarrow\mathcal L_{kp}^{\chi(\ell)}\mathbf h_i^{\star(\ell)}+\delta_{\ell0}\mu_{kp}^{\chi}$, $\chi\in\{\mathrm L,\mathrm R\}$.

\For{$g=1,\dots,N_{\rm q}$}
\State $\mathcal M_g\leftarrow\mathcal M(\mathbf r_g)$ and $\{\boldsymbol\delta_\eta,r_\eta,\widehat{\boldsymbol\delta}_\eta\}_{\eta\in\mathcal M_g}$ from Eq.~\ref{eq:query-images}.
\State $\phi_{\eta k_0}^{\chi}(\mathbf r_g),\,\Phi_k^{\chi}(\mathbf r_g)\leftarrow$ Eq.~\ref{eq:continuous-fields}.
\State $\rho_{\rm init}(\mathbf r_g)\leftarrow\displaystyle\sum_{\eta\in\mathcal M_g}\sum_{k_0}\phi_{\eta k_0}^{\mathrm L}(\mathbf r_g)\phi_{\eta k_0}^{\mathrm R}(\mathbf r_g)$, $\rho_{\rm env}(\mathbf r_g)\leftarrow\displaystyle\sum_k\Phi_k^{\mathrm L}(\mathbf r_g)\Phi_k^{\mathrm R}(\mathbf r_g)$.
\State $\widehat\rho_{\boldsymbol\theta}(\mathbf r_g;\mathcal X)\leftarrow\rho_{\rm init}(\mathbf r_g)+\rho_{\rm env}(\mathbf r_g)$.
\EndFor

\State \Return $\{\widehat\rho_{\boldsymbol\theta}(\mathbf r_g;\mathcal X)\}_{g=1}^{N_{\rm q}}$
\end{algorithmic}
\end{algorithm}

During training, the $\rho_{\rm init}$ branch is briefly pretrained with the remaining model parameters frozen, after which the complete model is jointly optimized using Eq.~\ref{eq:training-objective}.

\section{Supplementary results}

\subsection{Ablation studies}
\label{sec:ablation}
We have already examined the effect of the maximum angular order $L$ on the performance of AIDEN in Figs.~\ref{fig:bar-hse} and \ref{fig:pbe-qm9-loss}. Here, we further evaluate the contributions of individual architectural components and the rationale behind their composition.

We first remove GIE by retaining only the element-dependent scalar initialization and setting the higher-order initial features to zero. To evaluate RRA, we preserve the complete edge correlation operation in TECE but replace its learned multi-head attention weights with the parameter-free cutoff-normalized weights $a_e^{(0)}=f_{\rm c}(r_e)/\sum_{e'\to i} f_{\rm c}(r_{e'})$. This replacement preserves cutoff weighting and neighbor normalization while removing the learned query-key scores, radial phase, and head-dependent adaptive aggregation. We next assess the two physically motivated components of the density decoder. Specifically, we remove the one-center branch $\rho_{\rm init}$ and disable the learnable radial correction of the GTO basis by setting $\zeta_{\ell p}(r)=0$ in Eq.~\ref{eq:gaussian-radial}. Finally, to examine the ordering of the interaction layers while keeping their number and parameter count fixed, we compare the full ACE$\rightarrow$TECE architecture with the reversed TECE$\rightarrow$ACE arrangement.

Considering computational efficiency, we use a fixed random subset of 2,000 structures drawn from the original PBE training split. All variants are trained for 100,000 steps with $L=4$ using the same training subset and validation set. For all variants retaining $\rho_{\rm init}$, we load the same pretrained one-center checkpoint to ensure consistent initialization. The best validation NMAE is evaluated on a fixed subset of 256 structures from the original PBE validation split. The results are summarized in Tab.~\ref{tab:module-ablation}.

\begin{table}[h]
\centering
\caption{\textbf{Module ablation results.} Parameter counts and best validation NMAE $\varepsilon_{\rho}$ for models trained on the same 2,000-structure PBE subset. Relative changes are calculated with respect to the full ACE$\rightarrow$TECE model.}
\label{tab:module-ablation}
\small
\begin{tabular}{lccc}
\toprule
Ablation setting & Parameters $\downarrow$ & Best val $\varepsilon_{\rho}$ [\%] $\downarrow$ & Relative to Full [\%] $\downarrow$ \\
\midrule
Full ACE$\rightarrow$TECE & 4.459M & \textbf{1.13151} & - \\
TECE$\rightarrow$ACE & 4.459M & 1.14572 & +1.26 \\
w/o GIE & 4.393M & 1.14622 & +1.30 \\
w/o RRA & 4.205M & 1.16120 & +2.62 \\
fixed GTO & 4.454M & 1.21505 & +7.38 \\
w/o $\rho_{\rm init}$ & 4.457M & 1.53550 & +35.70 \\
\bottomrule
\end{tabular}
\end{table}

Removing GIE increases the best validation NMAE by 1.30\% relative to the full model, indicating that constructing higher-order equivariant initial representations from one-hop neighborhood geometry before the main interaction blocks facilitates the extraction of local directional information. Replacing RRA with fixed cutoff-normalized weights increases the NMAE by 2.62\%. Compared with assigning neighbor contributions solely according to distance, RRA combines equivariant query-key matching, radial rotary phases, and multi-head aggregation to adaptively select edge messages according to the local chemical and geometric environment. The decoder ablations lead to more pronounced performance degradation. Removing $\rho_{\rm init}$ increases the NMAE by 35.70\%, demonstrating the importance of the element-dependent one-center density as an atomic-density baseline, which allows the environment-dependent branch to focus on density corrections induced by bonding and coordination. Fixing the GTO basis increases the NMAE by 7.38\% while reducing only 4,590 parameters, indicating that the learnable correction $\zeta_{\ell p}(r)$ effectively adjusts the shapes of different angular and radial channels on top of the analytical even-tempered GTO basis to accommodate diverse elemental compositions and local bonding environments. The comparison of interaction order further supports the ACE$\rightarrow$TECE design. Reversing the full model to TECE$\rightarrow$ACE while retaining the same parameter count increases the NMAE by 1.26\%. This result suggests that first constructing atom-centered representations with cross-neighbor correlations through the aggregate-then-correlate operation of ACE, followed by directional refinement in edge-aligned local coordinate frames through the correlate-then-aggregate operation of TECE, provides a more effective ordering for charge-density learning.

Overall, these ablation results validate the effectiveness of GIE, RRA, one-center density initialization, and learnable GTO radial corrections, while also supporting the ordered ACE$\rightarrow$TECE interaction architecture. Among these components, $\rho_{\rm init}$ and the GTO correction provide the most substantial accuracy gains, whereas the ACE$\rightarrow$TECE ordering effectively combines cross-neighbor correlation modeling with edge-wise directional refinement.

\subsection{Training details}

\begin{figure}[h]
    \centering
    \includegraphics[width=\linewidth]{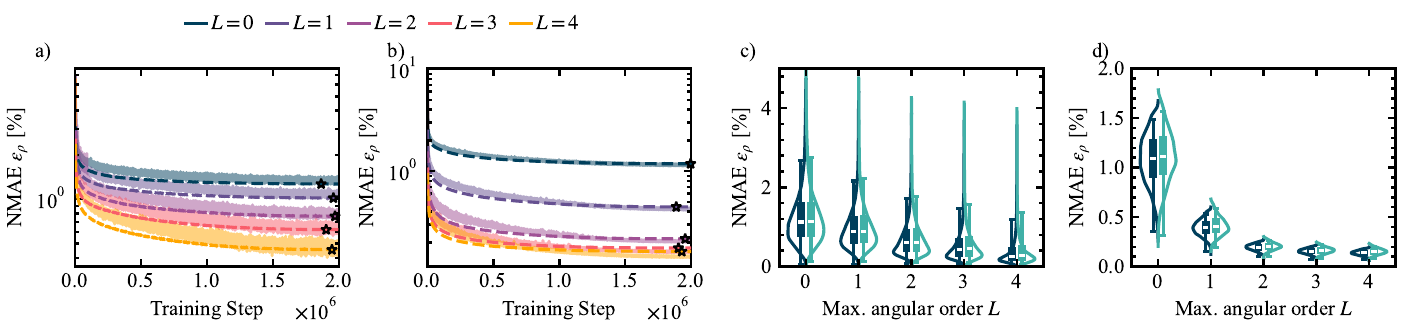}
    \caption{\textbf{AIDEN loss curves and violin plots of error distributions.} (a), (b) Loss curves of AIDEN on the PBE and QM9 datasets under different $L$ settings. Opaque dashed lines denote the validation set, and semi-transparent solid lines denote the training set. (c), (d) Violin plots of the error distributions of AIDEN on PBE and QM9. Dark blue and cyan respectively denote the training and validation errors. In the embedded box plots, white markers indicate the mean error, the filled boxes span the 25\%-75\% quantile range, and the upper and lower whiskers are set to the 5\% and 95\% quantiles.}
    \label{fig:pbe-qm9-loss}
\end{figure}

Fig.~\ref{fig:pbe-qm9-loss}(a-b) shows the loss curves of AIDEN on the PBE crystal dataset and the QM9 molecular dataset. For each dataset, we train models with maximum angular orders $L=0,\ldots,4$ while keeping all other training settings unchanged. The models are optimized using Adam, with learning rates of $2\times10^{-4}$ and $1\times10^{-3}$ for the encoder and decoder, respectively. We use a batch size of 12 and a gradient-clipping threshold of 0.5. Exponential moving average parameters with a decay rate of 0.995 are used for validation and inference. All models are trained on a single NVIDIA A100-40GB GPU.

\subsection{Twisted bilayer graphene}
\label{app:tbg-results}

\textbf{Structures and calculation settings.}
We consider three unrelaxed commensurate TBG cells, summarized in Tab.~\ref{tab:tbg-settings}. All structures have a 20~\AA\ out-of-plane lattice parameter. We reuse the converged PBE-SCF densities, energies, and forces as references and evaluate the complete model density grids without charge renormalization. For fixed-density calculations, the remaining VASP restart quantities and electronic settings are kept identical between AIDEN and ChargE3Net within each structure. We use the carbon PAW-PBE potential (08Apr2002), a 520~eV cutoff, Gaussian smearing of 0.05~eV, and non-spin-polarized calculations with fixed atomic positions and symmetry disabled. The fixed-density energy/force and band calculations use \texttt{ICHARG=11}, with electronic convergence thresholds of $10^{-7}$ and $10^{-8}$~eV, respectively, and \texttt{PREC=Accurate}, \texttt{LREAL=.FALSE.}, \texttt{LASPH=.TRUE.}, and \texttt{ADDGRID=.TRUE.}. Band calculations additionally use \texttt{LMAXMIX=2}. Energies are the final \texttt{TOTEN} values, and forces are the frozen-density approximate forces returned by VASP. The reciprocal-space path uses fractional coordinates $(0,0,0)\to(1/2,0,0)\to(1/3,1/3,0)\to(0,0,0)$. We use coarser line-mode sampling for the two larger cells to control memory use. The resulting comparisons contain 180, 18, and 18 $\mathbf{k}$ points, respectively, including repeated segment endpoints. It should be noted that the angular trend in the band errors is therefore evaluated at different sampling densities.

\begin{table}[h]
\centering
\caption{\textbf{TBG structures and sampling.} The $\Gamma$-centered mesh is used for the reference SCF and fixed-density energy/force calculations. The last two columns give the number of points per segment along $\Gamma$-M-K-$\Gamma$ and the number of returned bands. Sampling and band counts are matched across methods within each cell.}
\label{tab:tbg-settings}
\small
\setlength{\tabcolsep}{5pt}
\begin{tabular}{ccccccc}
\toprule
Twist angle & $N_{\rm a}$ & FFT grid & $N_{\rm g}$ & SCF mesh & Points/segment & Bands \\
\midrule
$21.79^\circ$ & 28 & $96\times96\times300$ & 2,764,800 & $9\times9\times1$ & 60 & 80 \\
$13.17^\circ$ & 76 & $160\times160\times300$ & 7,680,000 & $6\times6\times1$ & 6 & 168 \\
$9.43^\circ$ & 148 & $224\times224\times300$ & 15,052,800 & $4\times4\times1$ & 6 & 312 \\
\bottomrule
\end{tabular}
\end{table}

\begin{figure}[h]
\centering
\includegraphics[width=.8\linewidth]{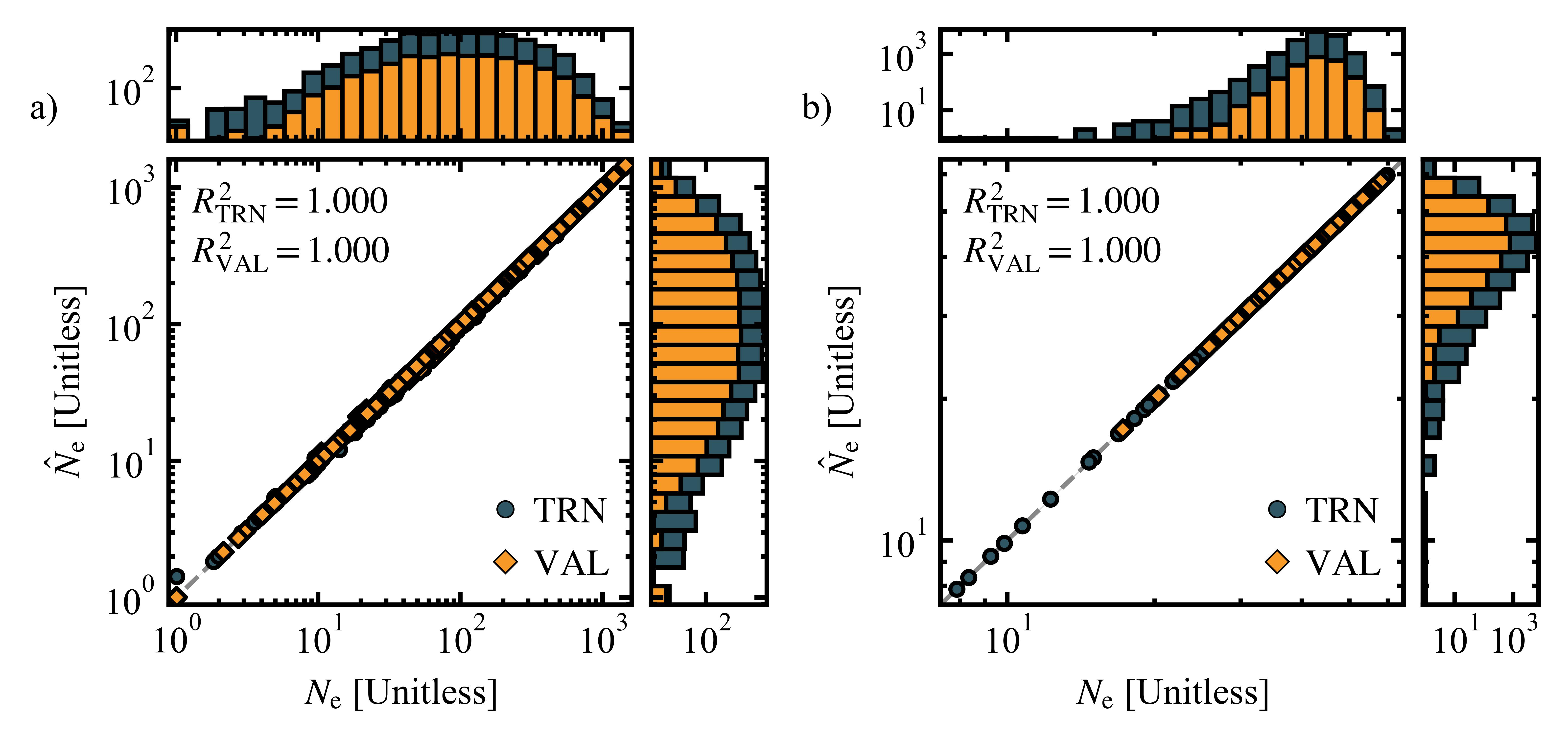}
\caption{\textbf{Parity plots of total charge on the general benchmark datasets.} Blue and yellow scatter points denote the training and validation samples in (a) PBE and (b) QM9, respectively. The horizontal axis $N_{\rm e}$ is the total DFT reference charge of each sample evaluated on the full grid, while the vertical axis $\hat{N}_{\rm e}$ is the total charge predicted by AIDEN, obtained by summing the charge density over all grid points. The marginal histograms along the horizontal and vertical axes show the numerical distributions of the total charge.}
\label{fig:charge-cons}
\end{figure}

\textbf{Charge density and charge conservation.}
\label{app:charge-count}
Tab.~\ref{tab:tbg-density} reports $\varepsilon_{\rho}$ using Eq.~\ref{eq:nmae}, together with the integrated smooth-grid electron counts. We denote the deviation of the predicted count from the reference by $\Delta N_{\rm e}$. AIDEN gives lower density errors at all three angles, with $\varepsilon_{\rho}=0.2892$-$0.2934\%$, compared with 0.3134-0.3160\% for ChargE3Net. Both models slightly overestimate the electron count, while AIDEN has the smaller deviation in every cell. Its relative charge error decreases from 0.00952\% to 0.00679\% as the cell grows, despite the increase in the absolute count deviation. Although we do not explicitly include a charge-conservation loss term in the training objective in Eq.~\ref{eq:training-objective} (as is also common in most related works), AIDEN naturally exhibits near-conservation of the total charge. This behavior is observed not only for the TBG cases but also across the general benchmark datasets (see Fig.~\ref{fig:charge-cons}), and can be largely attributed to the low overall error in charge density prediction.

\begin{figure}[t]
\centering
\includegraphics[width=\linewidth]{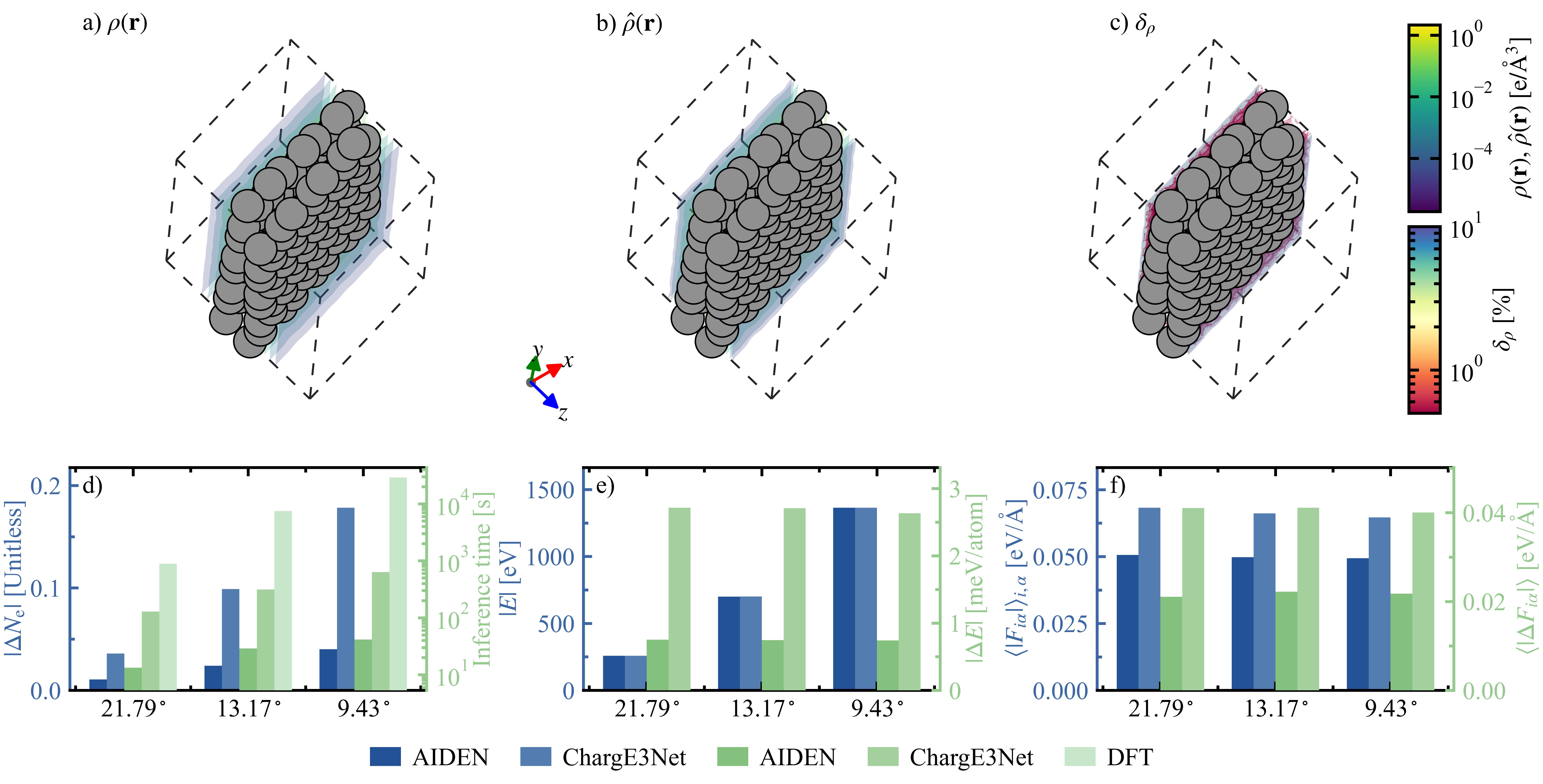}
\caption{\textbf{Real-space density and property comparison for TBG.} (a), (b) DFT and AIDEN density isosurfaces, $\rho(\mathbf r)$ and $\hat{\rho}(\mathbf r)$, shown with a common logarithmic color scale and identical isovalues. (c) Pointwise relative density error $\delta_{\rho}$, with isosurfaces at 0.5\%, 5\%, and 10\%. (d) Absolute electron-count deviation $|\Delta N_{\rm e}|$ (left axis) and full-grid density inference time (right axis). (e) Total-energy magnitude (left axis) and per-atom energy error $|\Delta E|$ (right axis). (f) Mean absolute force component $\langle|F_{i\alpha}|\rangle$ (left axis) and force-component MAE $\langle|\Delta F_{i\alpha}|\rangle$ (right axis). Subplots (d)-(f) compare AIDEN and ChargE3Net across all three twist angles; blue and green shades correspond to the left and right axes, respectively.}
\label{fig:tbg-all}
\end{figure}

For fixed learned species profiles, integrating the periodic one-center sum gives an extensive contribution $\int_\Omega\rho_{\rm init}(\mathbf r)\,{\rm d}^3\mathbf r=\sum_i q_{Z_i}$, where $q_Z$ is the integral of one localized species profile. This branch can therefore supply an atomic baseline for the electron count without learning a global constraint. The remaining environment-dependent contribution is not constrained to have zero integral, and the present results do not isolate the contributions of the two branches. Fig.~\ref{fig:tbg-all}(a)-(c) complements the integrated errors with a spatial comparison of the reference and predicted densities. The density panels use four logarithmically spaced isovalues within their shared positive density range. The pointwise error in panel (c) is expressed relative to the local reference density, with the denominator bounded below by $10^{-12}$ times its maximum magnitude to stabilize the low-density region. Panel (d) then relates the integrated charge deviation to the cost of evaluating the complete density field.

\begin{table}[t]
\centering
\caption{\textbf{Density and charge conservation for TBG.} Electron counts and their signed deviations are reported in units of $e$; relative deviations use the matched reference count.}
\label{tab:tbg-density}
\small
\setlength{\tabcolsep}{5pt}
\begin{tabular}{clccccc}
\toprule
Twist angle & Method & $\varepsilon_{\rho}$ [\%] & Reference count & Predicted count & $\Delta N_{\rm e}$ & Relative [\%] \\
\midrule
\multirow{2}{*}{$21.79^\circ$} & AIDEN & 0.293371 & 112 & 112.010667 & +0.010667 & 0.009524 \\
& ChargE3Net & 0.315961 & 112 & 112.035903 & +0.035903 & 0.032056 \\
\midrule
\multirow{2}{*}{$13.17^\circ$} & AIDEN & 0.290678 & 304 & 304.024022 & +0.024022 & 0.007902 \\
& ChargE3Net & 0.314475 & 304 & 304.098970 & +0.098970 & 0.032556 \\
\midrule
\multirow{2}{*}{$9.43^\circ$} & AIDEN & 0.289184 & 592 & 592.040188 & +0.040188 & 0.006788 \\
& ChargE3Net & 0.313427 & 592 & 592.178324 & +0.178324 & 0.030122 \\
\bottomrule
\end{tabular}
\end{table}

\textbf{Energies and forces.}
Tabs.~\ref{tab:tbg-energy} and \ref{tab:tbg-forces} give the numerical results corresponding to Fig.~\ref{fig:tbg-all}(e), (f). AIDEN gives $|\Delta E|=0.753$, 0.747, and 0.741~meV/atom, approximately 28\% of the corresponding ChargE3Net errors. The signed total-energy differences are negative for both models at every angle. The force-component MAEs remain approximately 0.021-0.022~eV/\AA\ for AIDEN and 0.040-0.041~eV/\AA\ for ChargE3Net. AIDEN also reduces the component RMSE and the maximum per-atom force-vector error in all three cells. These results show that its improvement in density accuracy is accompanied by consistent improvements in the fixed-density energy and force estimates.

\begin{table}[t]
\centering
\caption{\textbf{TBG total energies and energy errors.} The DFT row gives the reused SCF reference. The signed difference $\hat E-E$ is in eV, while $|\Delta E|$ follows the per-atom definition in Tab.~\ref{lab:ood-bench}.}
\label{tab:tbg-energy}
\small
\begin{tabular}{cllll}
\toprule
Twist angle & Method & Total energy [eV] & $\hat E-E$ [eV] & $|\Delta E|$ [meV/atom] \\
\midrule
\multirow{3}{*}{$21.79^\circ$} & DFT & -258.04561259 & 0 & 0 \\
& AIDEN & -258.06669840 & -0.02108581 & 0.753065 \\
& ChargE3Net & -258.12164598 & -0.07603339 & 2.715478 \\
\midrule
\multirow{3}{*}{$13.17^\circ$} & DFT & -700.39986467 & 0 & 0 \\
& AIDEN & -700.45662887 & -0.05676420 & 0.746897 \\
& ChargE3Net & -700.60567255 & -0.20580788 & 2.707998 \\
\midrule
\multirow{3}{*}{$9.43^\circ$} & DFT & -1363.97698248 & 0 & 0 \\
& AIDEN & -1364.08670932 & -0.10972684 & 0.741398 \\
& ChargE3Net & -1364.36657021 & -0.38958773 & 2.632350 \\
\bottomrule
\end{tabular}
\end{table}

\begin{table}[t]
\centering
\caption{\textbf{TBG force magnitudes and errors.} All entries are in eV/\AA. The component RMSE averages squared errors over all atoms and Cartesian components; the last column is the maximum Euclidean norm of the force error over atoms.}
\label{tab:tbg-forces}
\small
\begin{tabular}{clcccc}
\toprule
Twist angle & Method & $\langle|F_{i\alpha}|\rangle$ & $\langle|\Delta F_{i\alpha}|\rangle$ & Component RMSE & Max. vector error \\
\midrule
\multirow{3}{*}{$21.79^\circ$} & DFT & 0.030418 & 0 & 0 & 0 \\
& AIDEN & 0.050602 & 0.021058 & 0.034872 & 0.062180 \\
& ChargE3Net & 0.068266 & 0.041057 & 0.062076 & 0.114601 \\
\midrule
\multirow{3}{*}{$13.17^\circ$} & DFT & 0.029764 & 0 & 0 & 0 \\
& AIDEN & 0.049782 & 0.022210 & 0.034972 & 0.065953 \\
& ChargE3Net & 0.066146 & 0.041128 & 0.061317 & 0.114324 \\
\midrule
\multirow{3}{*}{$9.43^\circ$} & DFT & 0.029303 & 0 & 0 & 0 \\
& AIDEN & 0.049377 & 0.021770 & 0.034843 & 0.067280 \\
& ChargE3Net & 0.064642 & 0.040043 & 0.060362 & 0.113618 \\
\bottomrule
\end{tabular}
\end{table}

\textbf{Band structures.}
For the band metrics in Tab.~\ref{tab:tbg-bands}, we compare equal band indices at matched $\mathbf{k}$ points and retain reference states within $E_{\rm Fermi}\pm5$~eV. This gives 4,450, 996, and 1,663 states for the three cells. Raw errors compare the eigenvalues directly; aligned errors apply one least-squares rigid shift per model and cell before computing the residuals. For the retained state set $\mathcal S$, the shift added to the model eigenvalues is $s=|\mathcal S|^{-1}\sum_{(n,\mathbf k)\in\mathcal S}(\epsilon_{n\mathbf k}-\hat\epsilon_{n\mathbf k})$. Aligned errors use $\hat\epsilon_{n\mathbf k}+s-\epsilon_{n\mathbf k}$, measuring residual band shape after removal of one global offset. Raw errors retain the offset under the calculation's energy-reference convention; they do not establish vacuum-referenced absolute levels. The DFT band calculation supplies $E_{\rm Fermi}$. Fig.~\ref{fig:tbg-band} presents the spectra around this reference.

\begin{table}[t]
\centering
\caption{\textbf{TBG band errors within $E_{\rm Fermi}\pm5$~eV.} All numerical entries are in eV. The shift is added to the model eigenvalues. MAE, RMSE, and maximum absolute error are evaluated over the same reference-state mask.}
\label{tab:tbg-bands}
\small
\setlength{\tabcolsep}{4pt}
\begin{tabular}{clrrrrrr}
\toprule
& & & \multicolumn{2}{c}{Raw} & \multicolumn{3}{c}{Aligned} \\
\cmidrule(lr){4-5}\cmidrule(lr){6-8}
Twist angle & Method & Shift & MAE & RMSE & MAE & RMSE & Max. abs. \\
\midrule
\multirow{2}{*}{$21.79^\circ$} & AIDEN & -0.179315 & 0.179315 & 0.179863 & 0.008415 & 0.014021 & 0.077684 \\
& ChargE3Net & -0.318982 & 0.319104 & 0.324463 & 0.027352 & 0.059382 & 0.327034 \\
\midrule
\multirow{2}{*}{$13.17^\circ$} & AIDEN & -0.172100 & 0.172100 & 0.173577 & 0.007410 & 0.022592 & 0.512770 \\
& ChargE3Net & -0.333283 & 0.333283 & 0.334107 & 0.010303 & 0.023454 & 0.294968 \\
\midrule
\multirow{2}{*}{$9.43^\circ$} & AIDEN & -0.168169 & 0.171246 & 0.184834 & 0.011468 & 0.076698 & 1.450108 \\
& ChargE3Net & -0.319113 & 0.323720 & 0.325494 & 0.012928 & 0.064134 & 1.437400 \\
\bottomrule
\end{tabular}
\end{table}

AIDEN has the lower aligned band MAE at every angle. Its raw band MAEs are 171-179~meV (Tab.~\ref{tab:tbg-bands}), with the rigid shift removing the dominant reference-energy offset. Its largest reduction in band MAE occurs at $21.79^\circ$, from 27.35 to 8.42~meV. Interestingly, the ranking of the tail errors is less uniform: ChargE3Net gives a smaller maximum aligned error at $13.17^\circ$, and a smaller aligned RMSE and maximum error at $9.43^\circ$. Both models reach maximum individual-state errors of approximately 1.44-1.45~eV in the largest cell.

\textbf{Density inference efficiency.}
Tab.~\ref{tab:tbg-inference} lists the full-grid density inference times used in Fig.~\ref{fig:tbg-all}(d). Both model timings are measured on a single NVIDIA A100 GPU after warm-up, with CUDA synchronization and model loading excluded; ChargE3Net uses its official full-grid per-sample inference pipeline. AIDEN performs the atomic encoding once per structure and reuses it across grid queries. Its inference time increases from 13.27 to 41.43~s as $N_{\rm g}$ increases from 2.76 to 15.05 million, whereas ChargE3Net requires 128.81-634.14~s. The measured speedups are 9.71, 10.85, and 15.31$\times$, respectively. The reported speedups refer to these full-grid implementations on the specified hardware.

\begin{table}[t]
\centering
\caption{\textbf{Full-grid charge density inference times for TBG.} Times refer to AIDEN's full-grid inference and ChargE3Net's official full-grid per-sample inference. The speedup is the ratio of ChargE3Net to AIDEN time.}
\label{tab:tbg-inference}
\small
\begin{tabular}{cccc}
\toprule
Twist angle & AIDEN [s] & ChargE3Net [s] & Speedup \\
\midrule
$21.79^\circ$ & 13.270 & 128.813 & $9.71\times$ \\
$13.17^\circ$ & 28.933 & 314.058 & $10.85\times$ \\
$9.43^\circ$ & 41.430 & 634.141 & $15.31\times$ \\
\bottomrule
\end{tabular}
\end{table}

\subsection{DFT calculations for water, AlMg, and a-Si}
\label{app:ood-settings}

\subsubsection{Structures and DFT settings}
We use the water and AlMg structures and one a-Si configuration from the OOD examples of Ref.~\citep{eacnet}, with $N_{\rm a}=192$, 108, and 64, respectively. The a-Si configuration is frame 11. The supplied files determine the structures and FFT grids; reference densities and properties are obtained from new PBE-SCF calculations. The density grids are $216^3$ for water, $180^3$ for AlMg, and $168^3$ for a-Si. All calculations use VASP with PAW-PBE potentials for H, O, Si, Al, and Mg, a 520~eV cutoff, and Gaussian smearing of 0.05~eV. We use non-spin-polarized calculations and hold atomic positions fixed. Water and a-Si use a $2\times2\times2$ Monkhorst-Pack mesh with symmetry disabled and an electronic convergence threshold of $10^{-8}$~eV, together with \texttt{PREC=Accurate}, \texttt{LREAL=.FALSE.}, \texttt{LASPH=.TRUE.}, \texttt{ADDGRID=.TRUE.}, and \texttt{LMAXMIX=2}. The recalculated AlMg benchmark uses \texttt{KSPACING=0.22}, yielding 14 irreducible $\mathbf{k}$ points, with \texttt{EDIFF=1E-4}, \texttt{ISYM=2}, \texttt{PREC=Accurate}, \texttt{LREAL=.FALSE.}, \texttt{LASPH=.FALSE.}, \texttt{ADDGRID=.FALSE.}, and \texttt{LMAXMIX=2}. Structure, potentials, grids, and electronic settings are matched across methods for each system.

\subsubsection{Density and property evaluation}
\label{app:property}
AIDEN uses its PBE-pretrained parameters, and ChargE3Net uses the public Materials Project pretrained checkpoint. SAD is generated with \texttt{ICHARG=12}. Each candidate density is subsequently evaluated with \texttt{ICHARG=11}, keeping the density fixed throughout electronic minimization. Model densities are used without charge renormalization. For AIDEN and ChargE3Net, the remaining VASP restart quantities are taken from the matched PBE-SCF reference and kept identical within each structure; SAD retains its own \texttt{ICHARG=12} initialization. Energies are the final VASP \texttt{TOTEN} values, and forces are the frozen-density approximate forces reported by VASP.

\subsubsection{Cross sections}
\label{app:cross-sec}
For Fig.~\ref{fig:ood-density}, candidate $x$-normal planes are the FFT planes nearest to atomic centers. We select the plane with the most atoms within 1.5 grid spacings, using proximity to the centers to break ties and periodic distances throughout. The selected zero-based indices are 191, 30, and 97 for water, AlMg, and a-Si, containing 9, 13, and 5 nearby atoms, respectively. AIDEN/ChargE3Net/SAD section NMAEs are 1.887/1.641/12.103\% for water, 1.183/1.648/11.405\% for AlMg, and 1.302/1.618/9.772\% for a-Si. The density panel and error panels use separate logarithmic color scales, with a common error scale across methods for each system. Display clipping is applied only when plotting.

\subsubsection{Timing}
Model inference is measured on a single NVIDIA A100 GPU after warm-up, with CUDA synchronization and model loading excluded. Both AIDEN and ChargE3Net use full-grid per-sample inference, with ChargE3Net evaluated through its official inference pipeline. VASP runs use 16 CPU MPI processes with one thread per process. The SAD time measures the complete \texttt{ICHARG=12} calculation, including diagonalization. The model-to-model speedup ratios compare the evaluated AIDEN and ChargE3Net full-grid implementations.

\subsubsection{Lower NMAE does not guarantee lower downstream errors}
\label{app:nmae-downstream}

The density NMAE and downstream errors measure different aspects of a prediction. For a single structure, Eq.~\ref{eq:nmae} integrates $|\delta\rho(\mathbf r)|$ with a uniform spatial weight, where $\delta\rho=\hat{\rho}-\rho$, and therefore ignores both the sign and spatial distribution of the error. In contrast, energy and force errors depend on where the density error occurs and how it is weighted by the electronic structure. For example, \cite{li2025image} shows that a lower density MAE does not necessarily lead to a smaller one-step DFT energy error. Similarly, density-driven exchange-correlation errors may follow a different ordering from density errors because of derivative dependence and cancellation between signed local contributions~\citep{10.1021/acs.jctc.7b00550}.

The local electron-ion interaction provides a simple illustration. For a fixed structure and local ionic potential $V_I^{\rm loc}$, the density-induced change in the corresponding energy contribution and the direct force contribution can be written as
\begin{align}
\delta E_{\rm loc}=\int_\Omega {\rm d}^3\mathbf r\,\delta\rho(\mathbf r)\sum_I V_I^{\rm loc}(\mathbf r-\mathbf R_I),
\label{eq:density-local-energy}\\
\Delta\mathbf F_I^{\rm loc}\simeq-\int_\Omega {\rm d}^3\mathbf r\,\delta\rho(\mathbf r)\frac{\partial V_I^{\rm loc}(\mathbf r-\mathbf R_I)}{\partial\mathbf R_I}.
\label{eq:density-local-force}
\end{align}
It should be noted that the force weight is a vector field, so density errors on different sides of an atom may reinforce or cancel according to their signs and directions. This information is not retained by either the full-grid NMAE or an integrated near-core absolute error. During fixed-density electronic minimization, the reported observables also include the orbital response to the supplied density; Eqs.~\ref{eq:density-local-energy} and \ref{eq:density-local-force} only isolate the local contribution. To further examine where the density error occurs, we integrate its magnitude over grid points within $R=0.6$~\AA\ of the nearest atom, including periodic images:
\begin{align}
D_R=\int_{\Omega_R}{\rm d}^3\mathbf r\,|\delta\rho(\mathbf r)|,\quad \Omega_R=\{\mathbf r\in\Omega:\min_{I,\mathbf n\in\mathbb Z^3}|\mathbf r-\mathbf R_I-\mathbf n\mathbf A|<R\}.
\label{eq:near-core-density-error}
\end{align}
For a-Si, $D_R$ is 0.1841~$e$ for AIDEN and 0.2437~$e$ for ChargE3Net, corresponding to a reduction of approximately 24\%. This is accompanied by a smaller AIDEN energy error, 3.213 versus 4.774~meV/atom, while the force-component MAEs are nearly identical, 0.04087 versus 0.04090~eV/\AA. For AlMg, $D_R$ is 0.2285~$e$ for AIDEN and 0.1520~$e$ for ChargE3Net, and ChargE3Net also has the smaller maximum pointwise error, 0.01001 versus 0.02456~$e$/\AA$^3$. Nevertheless, AIDEN achieves the lower global density NMAE. These results again show that different density metrics emphasize different spatial characteristics and need not induce the same ordering as downstream energy or force errors. More generally, previous work has also shown that spatially localized density inaccuracies can contribute disproportionately to electron-nuclear energy errors~\citep{lewis2021learning}. Thus, $D_R$ is useful as a spatial diagnostic, but it should not be interpreted as a direct predictor of the signed energy or force response. Water provides another clear example. ChargE3Net has the smaller $D_R$, 4.1411 versus 4.5969~$e$, whereas AIDEN yields the smaller total force-component MAE. Resolving the forces by element gives 0.17663/0.17008~eV/\AA\ for O and 0.08370/0.10303~eV/\AA\ for H, in AIDEN/ChargE3Net order. Since the cell contains 64 O and 128 H atoms, the overall MAE is $(\mathrm{MAE}_{\rm O}+2\mathrm{MAE}_{\rm H})/3$. In our case, the improvement on H is sufficiently large to outweigh the slightly larger O error, giving an overall MAE of 0.11467~eV/\AA\ for AIDEN compared with 0.12538~eV/\AA\ for ChargE3Net.

One possible interpretation is that AIDEN better captures the signed and directional density redistribution that is relevant to H forces. In water, local polarization and charge redistribution vary with the surrounding hydrogen-bond environment~\citep{10.1021/jp067922u,10.1063/1.4990504}. Such directional changes can strongly affect the vector-weighted force response without necessarily reducing an integrated absolute density error. We therefore regard the element-resolved force results as direct evidence that the two models distribute their density errors differently, while the connection to local polarization remains a physical interpretation rather than a separately verified mechanism. The total-energy ranking also depends on cancellations among different energy contributions. \cite{li2025image} reports individual component errors that are substantially larger than the final total-energy error, highlighting the importance of such cancellation. For water, the signed errors $\hat E-E$ are $+0.66117$~eV for AIDEN and $-1.23700$~eV for ChargE3Net, placing the two predictions on opposite sides of the reference. Therefore, the smaller AIDEN absolute error does not imply that every individual energy contribution is more accurate. Moreover, for sufficiently small, electron-number-conserving perturbations around self-consistency, the stationary Harris-Foulkes construction cancels first-order energy variations, leaving a leading quadratic response determined by both the perturbation and the electronic response~\citep{10.1103/PhysRevB.39.12520}. This provides another reason why a scalar density norm cannot be expected to impose a monotonic ordering on total-energy errors. Overall, our results show that global density accuracy, spatially resolved density errors, and downstream observables provide complementary rather than interchangeable assessments of learned electron densities.

\end{document}